\documentclass[aps,pra,floatfix,twocolumn,superscriptaddress,10pt,balancelastpage]{revtex4-1}
\usepackage{amssymb, amsmath,graphicx,subfigure}
\usepackage{color}
\usepackage{braket}
\usepackage{hyperref}
\usepackage{times}
\usepackage{booktabs}

\begin{document}

\title{Persistent-mediator architecture for universal fault-tolerant quantum computation with color codes}

\author{Zhi-Cheng He}
\affiliation{State Key Laboratory of Low Dimensional Quantum Physics, Department of Physics, Tsinghua University, Beijing, 100084, China}
\affiliation{Frontier Science Center for Quantum Information, Beijing 100184, China}

\author{Jingbo Wang}
\affiliation{Beijing Academy of Quantum Information Sciences, Beijing 100193, China}

\author{Dong E. Liu}
\thanks{Corresponding author. Email: dongeliu@mail.tsinghua.edu.cn}
\affiliation{State Key Laboratory of Low Dimensional Quantum Physics, Department of Physics, Tsinghua University, Beijing, 100084, China}
\affiliation{Frontier Science Center for Quantum Information, Beijing 100184, China}
\affiliation{Beijing Academy of Quantum Information Sciences, Beijing 100193, China}
\affiliation{Hefei National Laboratory, Hefei 230088, China}

\begin{abstract}
Two-dimensional color codes provide transversal Clifford gates, whereas suitable three-dimensional codes provide a transversal $T$ gate. Representative cross-dimensional schemes either switch the computational encoding or consume a 3D encoded state. Reusing the same protected 3D block would remove repeated auxiliary-state preparation and terminal readout, raising a central question: can the 3D module remain encoded and protected while mediating successive $T$ gates on a separate 2D register? Here we introduce a cross-dimensional persistent-mediator architecture that retains the 2D data register and 3D mediator as distinct encoded modules at each cycle boundary, coupled only through a transient protected subsystem. We construct a membrane-like joint-logical measurement and combine it with a matched cross-dimensional $CZ$, complementary string-like surgery, and feed-forward to close a deterministic logical $T$-gate cycle. The cycle avoids post-selection and destructive mediator readout and restores the 3D block to its initial logical state. Stabilizer analysis establishes the logical action and preserved spatial distance. Phenomenological and circuit-level simulations of a Clifford surrogate show finite-distance scaling close to the matched 3D memory benchmark, while sequential tests show approximately linear logical-error accumulation in the low-error regime. Persistent mediation enables repeated access to higher-dimensional non-Clifford capability while keeping the computational register in its 2D encoding.

\end{abstract}

\maketitle

\section{Introduction}

Fault-tolerant quantum computation requires an architecture to protect encoded information while supplying a universal set of logical gates. Quantum error correction suppresses physical faults at the encoded level~\cite{QEC_Shor,QEC_Gottesman,QEC_Steane,QEC_Knill}, and recent logical-processor experiments have made protected gate operations increasingly relevant~\cite{Surface_code_google,Single_qubit_gate_2,Bravyi_2024,atom_arrays}. Transversal gates are particularly attractive because they restrict error propagation, yet the Eastin--Knill theorem precludes a universal transversal gate set for a quantum error-correcting code~\cite{Eastin_universality}. Universal computation therefore requires combining complementary protected operations. The architectural challenge is to incorporate the missing gate capability without sacrificing fault-tolerant protection or repeatedly rebuilding the encoded resource that provides that capability.

Color codes make the tension particularly clear. Triangular two-dimensional (2D) color codes support transversal Clifford operations but lack the transversal non-Clifford $T$ gate needed to complete the standard universal gate set; suitable tetrahedral three-dimensional (3D) color codes provide the complementary transversal $T$ operation~\cite{3D_color_1,3D_color_2}. Dimensional code switching accesses the 3D capability by changing the encoding while preserving logical information~\cite{code_switching,code_switching_2,code_switching_3,Beverland_Kubica_Svore_Cost}. Directed transversal controlled gates provide another route through teleportation-based protocols between matched 2D and 3D codes~\cite{Double_code,Double_code_2,code_switching_with_CNOT,code_switching_with_CNOT_2,code_switching_with_CNOT_3}. In both approaches, non-Clifford capability is supplied by the higher-dimensional code while the computational register is naturally maintained in 2D. The key difficulty is to connect the two encoded systems while preserving protection throughout the complete non-Clifford operation.

Lattice surgery provides a natural mechanism for protected coupling through joint logical measurements~\cite{Lattice_surgery_Horsman,Lattice_surgery_experiment,Lattice_surgery_0,Lattice_surgery_2}. Subsystem lattice surgery further established fault-tolerant interfaces between distinct encoded memories and processors, allowing logical information to be transferred between different code families~\cite{Lattice_surgery_0}. 
Generalized lattice-surgery teleportation can apply a transversal non-Clifford gate in a 3D auxiliary code and transfer the resulting logical action to a separate 2D computational register~\cite{Lattice_surgery_1}. In the representative implementations relevant here, destructive logical readout then consumes the auxiliary encoded state. The physically relevant distinction is whether the higher-dimensional block remains encoded, protected, and available for the next non-Clifford operation. The resulting question is sharp: can a complete fault-tolerant $T$-gate cycle act on a 2D color-code register while preserving the 3D block as an active encoded module and returning the mediator to a well-defined logical state for subsequent use?

A closed cycle requires more than the return of an ancilla to its initial state in an ideal logical circuit. A protected implementation requires extraction of the intended joint logical parities, preservation of the surviving logical degree of freedom and its spatial distance, a consistent syndrome history across the merge--split boundaries, and circuit-level extraction of high-weight checks for avoiding correlated-error propagation. The 3D color-code geometry makes these requirements nontrivial: one logical Pauli operator is naturally membrane-supported, whereas the complementary 2D logical operator admits a string-like representative. The key missing ingredient is therefore a protected cross-dimensional interface compatible with both geometries and capable of closing the logical cycle while retaining the 3D block.

Here we construct a cross-dimensional persistent-mediator architecture for universal fault-tolerant quantum computation with color codes. A protected heterogeneous interface closes the non-Clifford cycle while the 2D computational register and 3D mediator remain distinct encoded modules at the cycle boundaries. Unlike dimensional code switching~\cite{code_switching,code_switching_2,code_switching_3,Beverland_Kubica_Svore_Cost}, the computational register is not converted into a 3D encoding; the higher-dimensional block remains independently encoded and repeatedly corrected across successive operations. The key interface is a membrane-like lattice-surgery measurement that couples the two blocks through a transient protected subsystem.
We prove that its local measurements reveal no logical information beyond the intended joint parity and that the merged subsystem retains the designed spatial distance. Combined with the matched logical CZ, complementary string-like surgery, the transversal $T$ operation, and Clifford feed-forward, the construction closes a deterministic logical $T$-gate cycle on the original 2D register. After correction, the 3D block returns to its initial logical state without post-selection or destructive logical readout and can remain encoded for subsequent non-Clifford operations.

Fault tolerance is analyzed through the merge--split implementation. We derive the merged stabilizer structure and spatial distance, use repeated syndrome measurements for spacetime protection, and implement high-weight checks with flag-based extraction circuits whose records enter the decoder~\cite{Chao_Reichardt_Two_Extra_Qubits,Chamberland_Beverland_Flag_Arbitrary_Distance,Chao_Reichardt_Flag_Any_Stabilizer_Code,Prabhu_Reichardt_Fewer_Qubits}. Numerical tests use a Clifford surrogate that preserves the interface geometry, circuit locations, syndrome schedule, and decoder inputs while replacing the non-Clifford layer by a stabilizer-compatible counterpart. The simulations characterize the protected interface rather than the exact noisy $T$ channel. For $d=3,5,7$, phenomenological and circuit-level results show finite-distance scaling close to the matched 3D memory benchmark under the same decoder. Sequential simulations further show approximately linear logical-error accumulation in the low-error regime under repeated use of the same mediator.

The architecture also changes how logical information and encoded resources are organized. Code switching accesses the non-Clifford gate by changing the encoding of the computational register~\cite{code_switching,code_switching_2,code_switching_3,Beverland_Kubica_Svore_Cost}, whereas teleportation-based schemes considered use a higher-dimensional auxiliary block whose encoded state is ultimately consumed by logical readout~\cite{code_switching_with_CNOT,code_switching_with_CNOT_2,code_switching_with_CNOT_3,Lattice_surgery_1}. Here the 2D data register and 3D mediator remain distinct encoded modules, coupled only through a temporary protected subsystem during the cross-dimensional operation. This separation keeps the computational register in the 2D code at the boundaries of every cycle and removes repeated preparation and terminal readout of a 3D encoded block from the per-gate resource flow. Each logical $T$ gate still requires the transversal $T$ layer, protected joint measurements, syndrome extraction, and feed-forward, so the overall cost remains implementation dependent. The main advantage is architectural: the 3D code becomes a continuously protected module whose non-Clifford capability can be accessed repeatedly without changing the encoding of the computational register or consuming the mediator after each gate.

\begin{figure*}[t]
\centering\includegraphics[width=\textwidth]{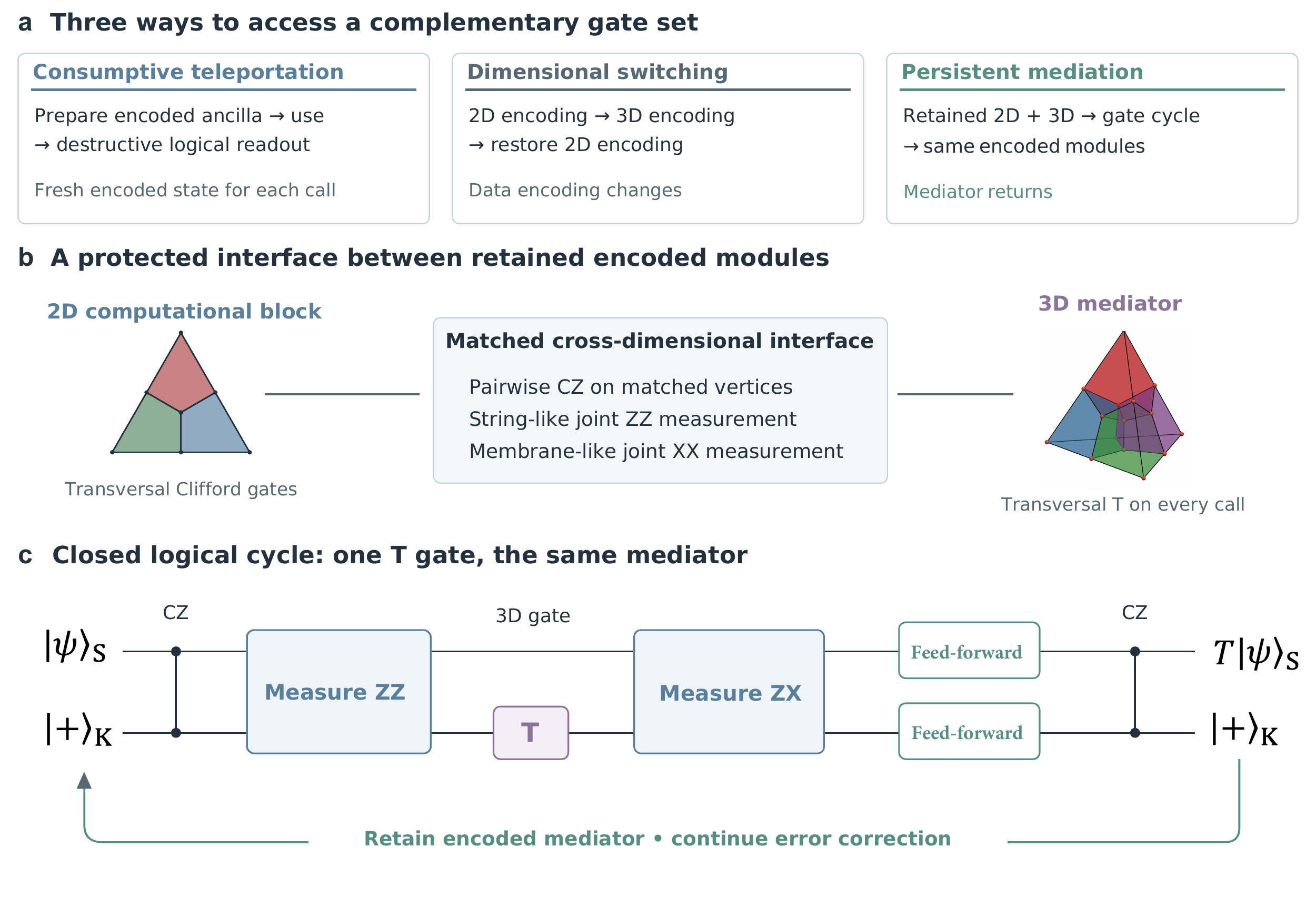}
\caption{\textbf{Persistent encoded mediation as an architectural operating mode.} \textbf{a}, Representative resource flows. Consumptive gate teleportation ends with destructive readout of an auxiliary encoded state; dimensional switching changes the data encoding; the present construction retains distinct 2D and 3D blocks across a closed cycle. \textbf{b}, A 2D computational block $S$ couples to a matched boundary membrane of a tetrahedral 3D mediator $K$. Pairwise controlled-$Z$ gates and complementary string- and membrane-like joint measurements implement the interface. \textbf{c}, Logical cycle, read from left to right. Starting from $\ket\psi_S\ket+_K$, the circuit applies $CZ$, measures $\bar Z_S\bar Z_K$, applies $T_K$, measures $\bar Z_S\bar X_K$, applies the outcome-dependent Clifford correction $C_{ab}$, and applies a final $CZ$. Here $a=m_{zz}$ and $b=m_{zx}$; the mixed measurement is implemented as $H_S$, $\bar X_S\bar X_K$ surgery, $H_S$. After correction, the output is $T_S\ket\psi_S\ket+_K$. The 3D encoded block is retained, but interface ancillas are reset and a new transversal $T_K$ layer is applied on every call.}
\label{fig:architecture}
\end{figure*}

\section{A closed logical cycle with a persistent mediator}

\begin{figure}[tbp]
\centering\includegraphics[width=\linewidth]{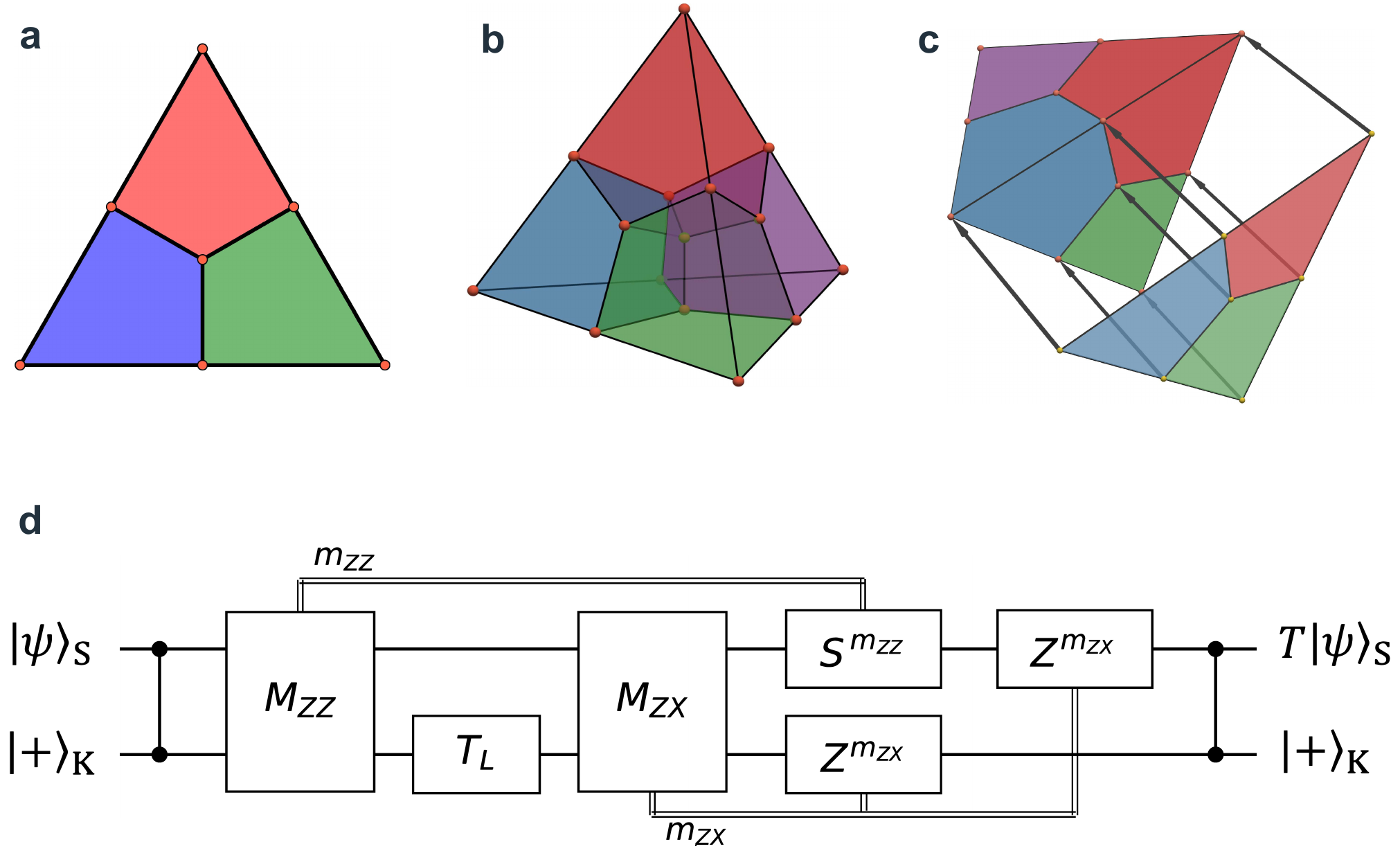}
\caption{\textbf{Code modules and logical implementation.} \textbf{a}, Distance-3 triangular 2D color code, with seven data qubits. \textbf{b}, Matched tetrahedral 3D color code, with fifteen data qubits. \textbf{c}, One-to-one controlled-$Z$ coupling between the 2D membrane and a matched 3D boundary membrane. Each physical qubit participates in at most one gate in this layer. \textbf{d}, Complete logical circuit, including the two joint measurements and outcome-dependent Clifford corrections. The final controlled-$Z$ disentangles the blocks and returns the mediator to $\ket+_K$.}
\label{fig:modules}
\end{figure}

The architecture assigns different gate capabilities to two encoded modules (Fig.~\ref{fig:architecture}). The 2D block $S$ provides the computational register and its transversal Clifford operations. The 3D block $K$ supplies its transversal $T$ operation through a matched interface. Standard preparation, readout and entangling Clifford operations between compatible 2D blocks complete the usual computational setting\cite{3D_color_1,3D_color_2,Lattice_surgery_0}; the construction below supplies the non-Clifford component.

For an arbitrary logical input $\ket\psi_S=\alpha\ket0_S+\beta\ket1_S$, initialize the mediator in $\ket+_K$. The cycle begins with a cross-dimensional logical $CZ$, followed by a trivial joint measurement of $\bar Z_S\bar Z_K$ with outcome $a\in\{0,1\}$, where the measured eigenvalue is $(-1)^a$. Apply the transversal $T_K$ layer, then measure $\bar Z_S\bar X_K$ with outcome $b\in\{0,1\}$. This second measurement uses a transversal Hadamard on $S$, a non-trivial $\bar X_S\bar X_K$ surgery, and a second Hadamard. The two surgery process would involve physical interface ancillas to update stabilizers, which we will discuss later. The cycle finally apply feed-forward operation
\begin{equation}
 C_{ab}=(S_S^{a}Z_S^{b})\otimes Z_K^{b}
 \label{eq:correction}
\end{equation}
and a second logical $CZ$. Here $S_S$ in the correction denotes the logical phase gate. The four corrections on logical qubit $(S,K)$ are $(I,I)$, $(Z,Z)$, $(S,I)$ and $(S^\dagger,Z)$, according to the measurement result $(a,b)=(0,0),(0,1),(1,0),(1,1)$, respectively.

Every corrected branch has the same ideal logical output,
\begin{equation}
 \ket\psi_S\ket+_K\longmapsto T_S\ket\psi_S\ket+_K.
 \label{eq:closedcycle}
\end{equation}
The branch derivation is given in Appendix, which retains the intermediate states. Since all the measurement outcomes are processed, there is no acceptance step in the logical cycle. During the cycle, the protected information can reside in correlations between the two blocks. Neither block is destructively read out, and the original 2D register again holds the output after disentangling. Therefore, the encoded mediator are reusable: it retaining the encoded mediator, while each call still applies a fresh transversal $T_K$ layer. 

The computational block $S$ and the mediator $K$ are protected by the repeated stabilizer measurements. When a logical $T$ gate is required, a logical cycle is applied to the corresponding codes. A completed call of logical cycle restores the separated-code check structure and the mediator's logical record, while not interrupt decoding of the space-time volume. The next call can then begin from the same ideal logical boundary condition. Notably, returning the ideal state to $\ket+_K$ does not remove accumulated physical faults. Therefore, error detection and correction of the mediator are nevertheless part of the logical cycle. Also, the classical measurement record is a part of the logical cycle, since an incorrect joint-parity estimate also causes an incorrect Clifford correction, even when the remaining logical operations are ideal. 

The logical mechanism of such a logical cycle is compatible with gate teleportation. Conjugation by $CZ$ maps $X_K$ to $Z_SX_K$; consequently, the mixed-parity readout is the corresponding non-destructive readout in an entangled frame. The contribution is the protected realization of that readout under the available color-code operations, together with a cycle that restores the encoded mediator. A replacement of destructive readout by a non-destructive measurement also needs to specify the necessary checks, auxiliary qubits or distance protection in the joint measurement process. We propose membrane-like surgery as the main interface to address those requirements.

\section{The membrane interface}

\begin{figure*}[tbp]
\centering\includegraphics[width=0.8\textwidth]{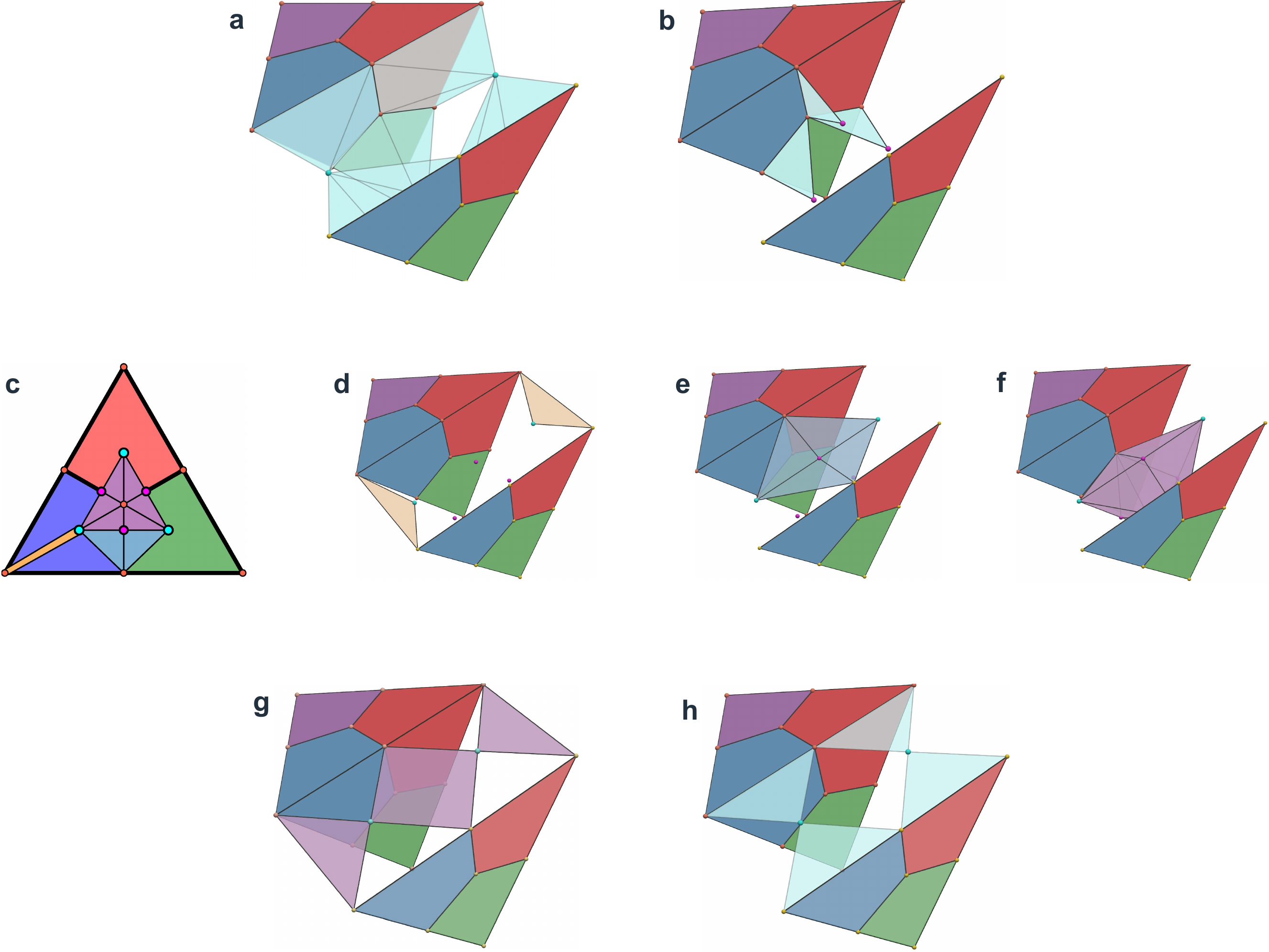}
\caption{\textbf{Complementary cross-dimensional surgery interfaces.} \textbf{a,b}, Modified $Z$ checks in the membrane interface: paired face checks share an A-type ancilla (cyan), and selected 3D checks extend through B-type edge ancillas (magenta). \textbf{c}, Top view of the new $X$ checks. \textbf{d--f}, Corner, boundary and bulk realizations of the single local rule in Eq.~\eqref{eq:interface}. The product of the checks is $\bar X_S\bar X_K$. \textbf{g,h}, Established string-like $\bar Z_S\bar Z_K$ surgery used for the complementary measurement: modified $X$ checks and new $Z$ checks, respectively. The membrane stage uses $\ket0$ interface ancillas, whereas the string stage uses $\ket+$ interface ancillas. Restoring the separated-code checks and measuring the interface ancillas completes each split. The diagrams show the distance-3 geometry.}
\label{fig:surgery}
\end{figure*}

We use the odd-distance triangular 6.6.6 color code and a tetrahedral color code with an isomorphic boundary facet (Fig.~\ref{fig:modules}), while the physical qubits occupy the vertices. The 2D stabilizers are face-supported $X$ and $Z$ operators; the 3D stabilizers are cell-supported $X$ and face-supported $Z$ operators. Write membrane $M_S$ for the full triangular patch of the 2D block $S$ and $M_K$ for the matched boundary membrane of 3D block $K$, with incidence-preserving vertex map $\pi_0$. Membrane-supported Pauli operators can be chosen as logical operators on both blocks. The even face weights and odd membrane size ensure their required stabilizer commutation and logical anti-commutation relations, see Methods for details.

Geometric matching does more than identify equal numbers of qubits. A 2D face must map to a 3D boundary face, and the intersection of a 3D cell with the membrane must pull back to a valid 2D check. These incidence conditions ensure that an $X$ stabilizer cannot propagate an unwanted logical $Z$ under the controlled gate. Therefore, for the vertexes $M_S^{(0)}$ in membrane, the the pairwise layer
\begin{equation}
 U_{CZ}=\prod_{v\in M_S^{(0)}}{CZ}_{v,\pi_0(v)}
 \label{eq:physicalcz}
\end{equation}
normalizes the product stabilizer group and induces a logical $CZ$ operation $\bar X_S\mapsto\bar X_S\bar Z_K$, $\bar X_K\mapsto\bar Z_S\bar X_K$, with both logical $Z$ operators unchanged. This logical action also follows by conjugating the established one-way $CNOT_{K\to S}$ with $H_S$\cite{Double_code,Double_code_2}. The smallest pair is the seven-qubit Steane code and the fifteen-qubit Reed--Muller code, while the same matched construction is used at the larger simulated distances.

The explicit matching is useful because it supplies the same geometry for the surgery construction. It  determines which ancillary qubits are needed when checks are extended during lattice surgery. The trivial lattice surgery protocol does not support 
the joint measurement of $\bar X_S\bar X_K$ since its 3D representative $\bar X_K$ is a membrane rather than a string. Thus we propose a membrane-like surgery to fault-tolerantly measure the parity. The construction should be designed to ensure any checks commute in the whole surgery process. Therefore, We modify and add stabilizer generators to establish the interface construction. We introduce interface ancillas on membrane faces and selected interior edges. An A-type face ancilla $a_f$ extends the paired 2D and 3D face checks, while a B-type edge ancilla $b_e$ extends a selected 3D face check whose boundary intersection is that edge. We initialize one ancilla $a_f$ on each membrane face and one $b_e$ on each selected interior edge in $\ket0$, and then modify the $Z$ checks near the membrane to cover the corresponding ancilla, shown in Figure~\ref{fig:surgery}.

For every matched vertex pair, we introduce the $X$-type check
\begin{equation}
 G_v=X_vX_{\pi_0(v)}\prod_{f\ni v}X_{a_f}\prod_{e\ni\pi_0(v)}X_{b_e}.
 \label{eq:interface}
\end{equation}
Whenever $G_v$ overlaps the data support of a modified $Z$ check, it also overlaps that check's ancilla. Thus, any checks in $G_v$ commute with modified $Z$ checks. Multiplying all $G_v$ cancels each face ancilla an even number of times and each edge ancilla twice:
\begin{equation}
 \prod_v G_v=X(M_S)X(M_K)\equiv\bar X_S\bar X_K.
 \label{eq:product}
\end{equation}
Thus local check outcomes determine the desired joint parity without destructive logical readout. Figure~\ref{fig:surgery} shows the resulting checks.

The ancillary initialization matters in this argument. Before the new $X$ checks are measured, each added physical qubit has a known $Z$ stabilizer, and the value of its corresponding $Z$ checks is known. Therefore, the value of modified $Z$ checks are deterministic. During the merge, these initialization constraints are replaced to be of modified $Z$ checks, while combinations that commute with the new measurements survive. Keeping those surviving constraints is essential when counting logical degrees of freedom.

For the complementary $\bar Z_S\bar Z_K$ measurement, we use established string-like surgery\cite{Lattice_surgery_0,Lattice_surgery_1,Lattice_surgery_2}. Its interface ancillas begin in $\ket+$; adjacent $X$ checks are extended and new $Z$ checks multiply to $\bar Z_S\bar Z_K$. The resulting checks are shown in Figure~\ref{fig:surgery}. The membrane construction could also support the complementary Pauli measurement, but the string interface uses fewer ancillas in this geometry. 

\section{Logical content and protection across the merge}
A commuting local check set is not, by itself, a characterization of the merged code. We therefore track the complete post-measurement stabilizer group, including inherited constraints. Before merging, the two blocks encode two logical qubits. The local measurements replace anticommuting constraints, while the joint logical parity adds one stabilizer $\bar X_S\bar X_K$. The merged state consequently retains one protected logical qubit, with representatives
\begin{equation}
 \bar X_M=[\bar X_S]=[\bar X_K],\qquad \bar Z_M=\bar Z_S\bar Z_K,
 \label{eq:mergedlogical}
\end{equation}
where brackets denote equivalence modulo the full post-measurement stabilizer group.

The spatial distance protection also survives the deformation. Restricting a merged logical $Z$ representative to either data block yields a nontrivial original logical $Z$ representative, giving $d_M^Z\geq d_S^Z+d_K^Z$. For the $X$ sector, multiplying merged stabilizers $G_v$ maps 3D stabilizers to 2D stabilizers, and maps the protected 3D logical $X$ class to the original 2D logical $X$ class without increasing weights. Conversely, the operation will also map 2D terms to 3D terms. Since a minimum-weight 2D logical $X$ remains valid after merging, we obtain $d_M^X=d_S^X=d$, and thus have $d_M=d$.
Measurement faults require a separate spacetime treatment. We repeat each merged-stage measurement for $r=d$ rounds and restore the separated-code checks at the split. The phenomenological argument uses initial and terminal detector constraints together with the spatial distance to obtain $d_M\geq\min(d,r)$ for the stated boundary conditions.

The split serves a different purpose from measuring the temporal stabilizers $G_v$. It removes the temporary interface degrees of freedom by directly measuring the ancilla, and restores the original block stabilizers, while preserving the post-measurement logical correlations required by the gate circuit. Tracking the corresponding measurement results, the separated-code syndrome history gives the decoder consistent information to detect errors.

At circuit level, high-weight extraction can propagate correlated hook errors. We use established flag-based measurement circuits\cite{Chao_Reichardt_Two_Extra_Qubits,Chamberland_Beverland_Flag_Arbitrary_Distance,Chao_Reichardt_Flag_Any_Stabilizer_Code,Prabhu_Reichardt_Fewer_Qubits}, to guarantee the error correction ability of all the code blocks. Any hook error can be detected by code-distance-dependent flag-based measurement circuits. We retain the flags as decoder inputs rather than rejecting flagged shots. The complete time-dependent circuit is evaluated numerically below.

\begin{figure*}[tbp]
\centering\includegraphics[width=0.8\textwidth]{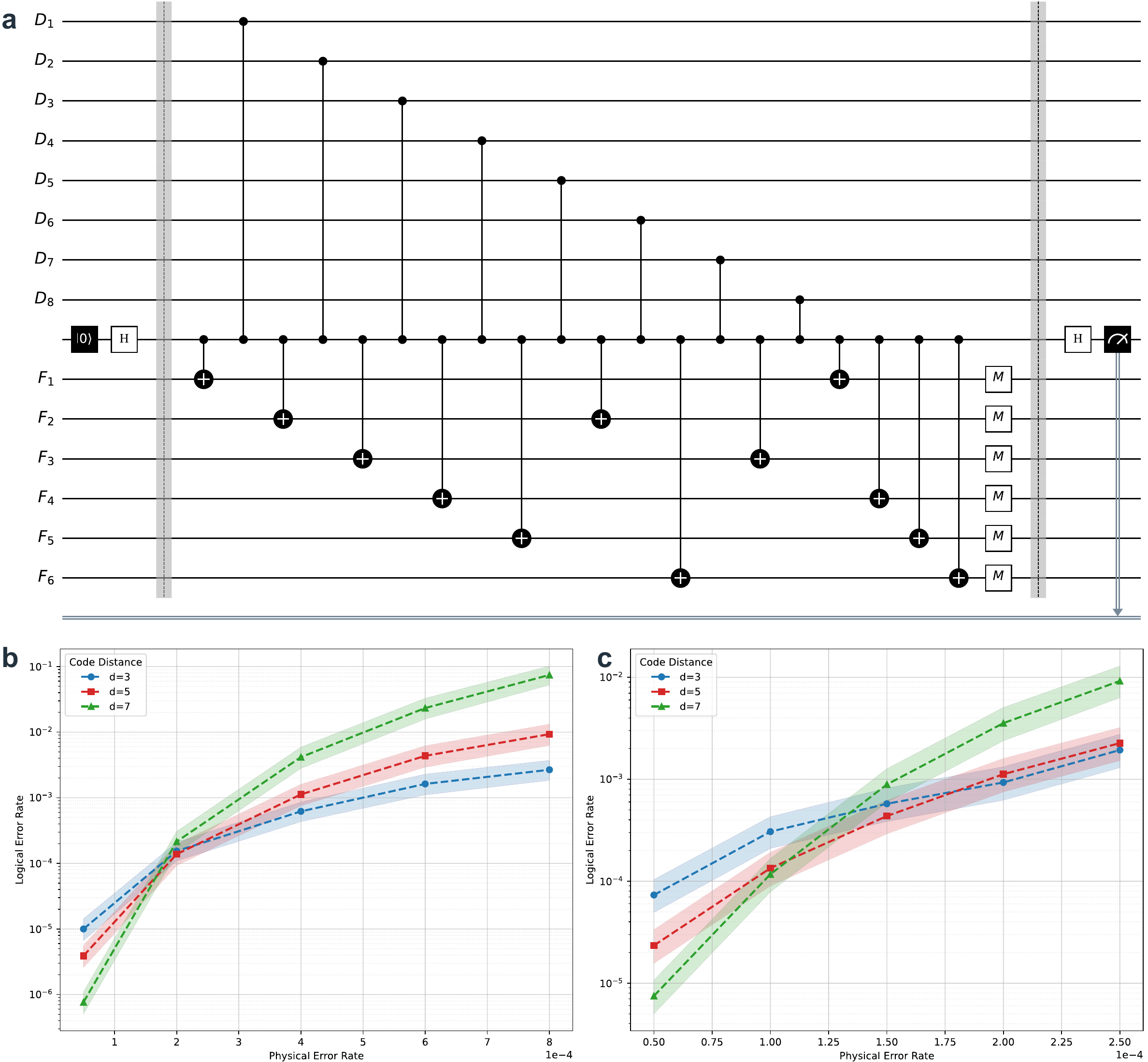}
\caption{\textbf{Flag-based extraction and circuit-level performance.} \textbf{a}, Weight-8 extraction circuit used in the distance-5 implementation; $D_i$ and $F_i$ label data and flag qubits. \textbf{b}, Matched 3D-memory benchmark. \textbf{c}, Clifford surrogate of the mediated interface. Both use the same single-parameter circuit-level noise model and hypergraph decoder, with all flag records retained. For $d=3,5,7$, the reported crossing regions are near $2\times10^{-4}$ and $1.5\times10^{-4}$, respectively.}
\label{fig:circuit}
\end{figure*}

\section{Performance and sequential reuse}

\begin{figure*}[tbp]
\centering\includegraphics[width=0.8\textwidth]{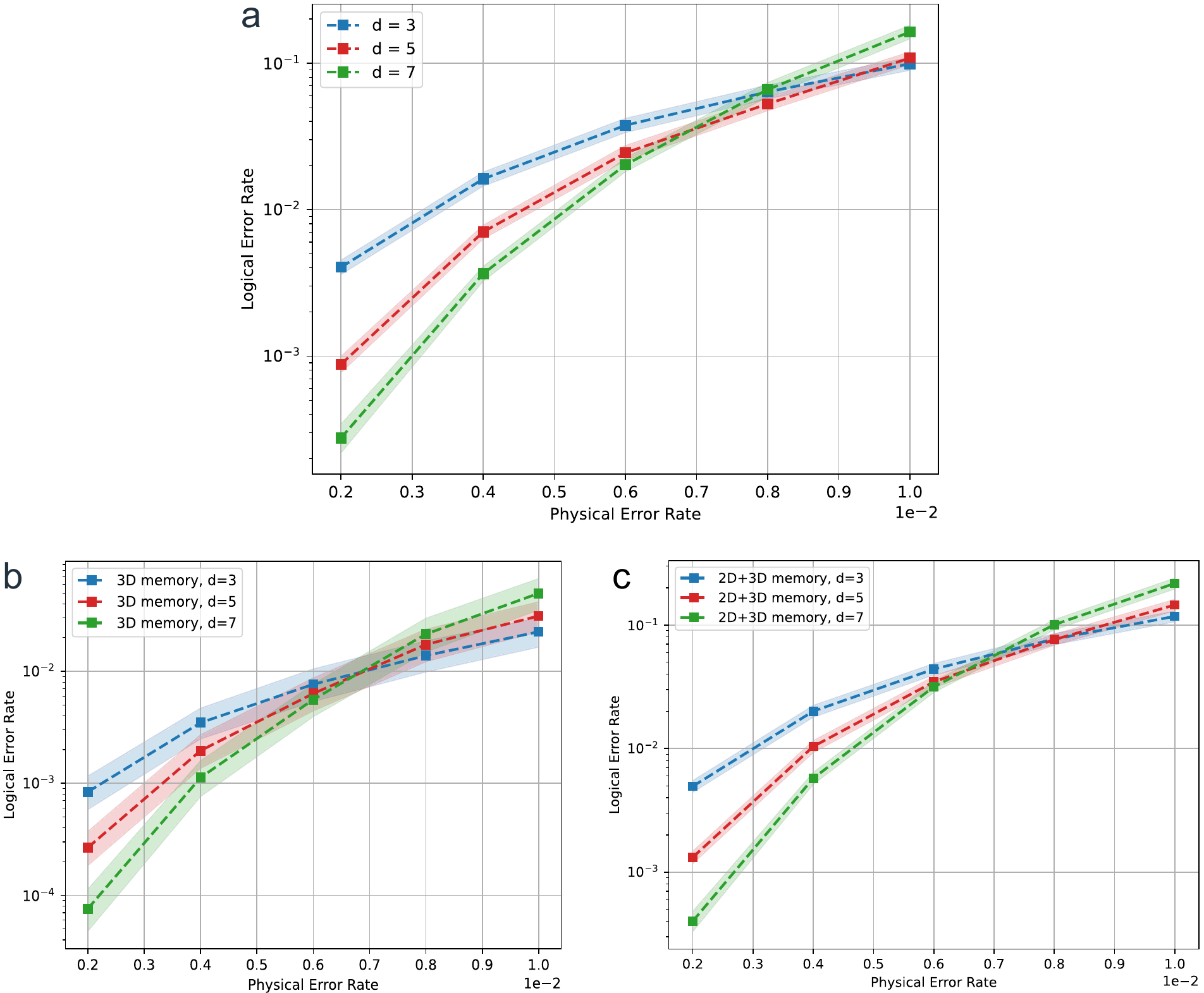}
\caption{\textbf{Phenomenological performance of the interface and memory simulations.} \textbf{a}, Clifford surrogate of the mediated gate at $d=3,5,7$, initialized in $\ket+_S\ket+_K$. The tracked binary observable is $y_S\oplus m_{zz}\oplus m_{zx}$, where $y_S$ is the final logical-$Y$ outcome. \textbf{b}, Matched 3D memory. \textbf{c}, Separated 2D-plus-3D memory. The memory simulations are initialized in the corresponding logical-$Y$ states using a noiseless logical $S$ operation and evolved for the same syndrome-round duration as the phenomenological interface test. Logical failure is a decoder disagreement with the expected observable. The crossing near $p\simeq0.7\%$ is finite-distance evidence under the stated hypergraph decoder.}
\label{fig:phen}
\end{figure*}

To test the encoded interface and the logical cycle at several distances, we replace each $T$ or $T^\dagger$ in the transversal pattern by $S$ or $S^\dagger$, respectively, and simulate the resulting Clifford surrogate with Stim and Sinter\cite{stim}. The surrogate preserves the geometry, interface operations and syndrome-extraction, but not exact fault propagation through a non-Clifford layer. A hypergraph union-find decoder processes the syndrome data\cite{hypergraphs_decoder,delfosse2021unionfinddecoderquantumldpc}. In every test, failure means disagreement between the decoder prediction and a specified deterministic logical observable.

Under phenomenological depolarizing noise, the surrogate shows distance-dependent error suppression and a common crossing region near $p\simeq0.7\%$ for $d=3,5,7$ (Fig.~\ref{fig:phen}). Matched 3D-only and separated 2D-plus-3D memory simulations use the corresponding encoding duration without the surgery operations. Their similar physical-error scale is consistent with the 3D component setting the dominant scale over the simulated distances. Since the simulation for higher code distance is not performed, we call these finite-distance crossings threshold evidence, rather than an asymptotic threshold estimate for the physical $T$ gadget.

\begin{figure}[tbp]
\centering\includegraphics[width=1.0\linewidth]{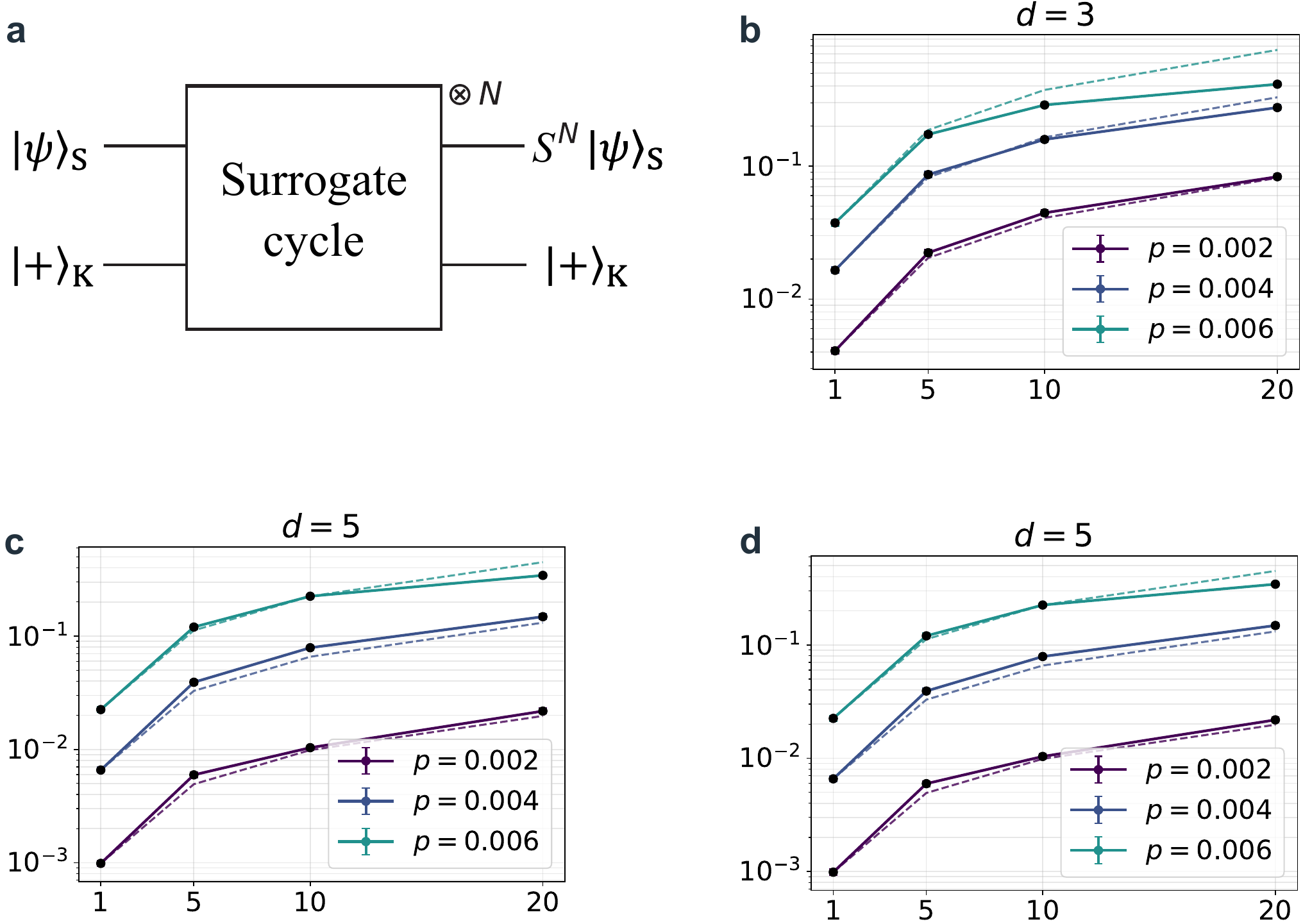}
\caption{\textbf{Sequential operation with a retained encoded mediator.} \textbf{a}, Repeated Clifford-surrogate cycle with a single initialization and final readout. Interface ancillas are reset between calls. \textbf{b--d}, Logical failure probability versus $N$ for $d=3,5,7$, respectively, at $p=0.002,0.004,0.006$. Solid curves show the simulation results; dashed curves are linear reference $NP_L(1)$. The logical observable is selected according to $S^N$. Near-linear accumulation is assessed only in the low-error regime; the logical error rate near one half corresponds to randomization of the measured logical qubit.}
\label{fig:reuse}
\end{figure}

Sequential tests retain the mediator through $N=1,5,10,20$ surrogate cycles, at $p=0.002,0.004,0.006$ (Fig.~\ref{fig:reuse}). In the sequential circuits, the same physical mediator participates in every call. These tests validate repeated operation of the implemented logical cycle while allowing for the incorporation of other transversal logical operations, which are not included in the current simulation. Initialization and terminal readout occur only once; interface ancillas are reset before each merge. The tracked logical observable follows the Clifford transformation $S^N$. For independent flips of that observable with probability $q$ per call, the reference expression is
\begin{equation}
 P_L(N)=\frac{1-(1-2q)^N}{2}\simeq Nq,\qquad Nq\ll1.
 \label{eq:reuse}
\end{equation}
In the low-error regime the accumulated failure probability is approximately linear in the number of calls. At larger error rates or longer sequences, saturation towards one half signals randomization of the tracked logical outcome.

Circuit-level tests include idle, single-qubit, two-qubit and measurement faults at a common probability $p$. The largest measured check has weight 24. The flag sequences and detector counts are retained in Appendix. We test the explicit flag-based interface implementation under the stated decoder and noise model. Figure~\ref{fig:circuit} shows crossing regions near $2\times10^{-4}$ for the 3D memory and $1.5\times10^{-4}$ for the surrogate of logical cycle. The two scales are comparable, but they are not identical. Considering large size of flag-based circuits, the sampling stops after at least 100 failures or one million shots, and uncertainty $21.6\%$ is reported with two-sided 95\% Wilson intervals. The complete circuit sizes also make the implementation cost visible. For distances three, five and seven, the largest number of flags in a single check circuit is six, eleven and twenty-five, respectively. The surrogate contains 1,718, 12,093 and 56,199 detectors, including both syndrome and flag information.

\section{Resource consequences}

\begin{table}[]
\centering
\small
\caption{Nominal logical-cycle resources.}\label{overhead}
\begin{tabular}{lcccc}
\hline
Quantity & \(d=3\) & \(d=5\) & \(d=7\) & General \(d\)\\
\hline
2D data qubits & 7&19&37&\((3d^2+1)/4\)\\
3D data qubits &15&65&175&\((d^3+d)/2\)\\
Membrane interface ancillas&6&24&54&\(3(d-1)^2/2\)\\
New interface \(X\) checks&7&19&37&\((3d^2+1)/4\)\\
Physical \(CZ\) gates per layer&7&19&37&\((3d^2+1)/4\)\\
Nominal stabilizer rounds&12&20&28&\(4d\)\\
Qubit-round volume&288&1820&6356&\((4d^4+9d^3+3d)/2\)\\
\hline
\end{tabular}
\end{table}

The architecture we proposed changes which operations recur in logical $T$ operations. A fresh 3D encoded state need not be prepared and destructively read out for every logical $T$; instead, the same block undergoes protected interactions, correction and frame update. However, we emphasize that the architecture removes such stage, but it does not remove the cost of keeping the 3D code active. We report the resource overhead of the logical-cycle in the architecture in the Table \ref{overhead}. It summarizes the resource counts. The $CZ$ gate count is given per layer, with two such layers per cycle. The interface count refers to the membrane stage. The qubit-round volume is obtained by summing over active data and interface qubits within each stage. Syndrome and flag ancillas, as well as extraction-gadget gate counts, are not included.The resource accounting uses $n_S=(3d^2+1)/4$ 2D data qubits, $n_K=(d^3+d)/2$ 3D data qubits and $n_I=3(d-1)^2/2$ membrane-interface ancillas. Each controlled-$Z$ layer contains $n_S$ physical gates, and the cycle contains two such layers. The nominal cycle contains $4d$ stabilizer rounds. Under this model the cycle scales as $O(d^4)$, as can deterministic preparation-based approaches involving an $O(d^3)$ 3D block and $O(d)$ verification rounds, leading to the same scaling $O(d^4)$ \cite{Beverland_Kubica_Svore_Cost,code_switching_with_CNOT_3}. An offline parallel preparation may reduce the latency of consumptive schemes, while a persistent mediator occupies protected hardware between calls.

Another relevant comparison is with a gate-teleportation cycle in which destructive auxiliary readout is replaced by a different protected non-destructive measurement. At the logical level that replacement can also permit a return condition, but its physical implementation still requires a fault-tolerant measurement procedure. The present membrane construction supplies one such procedure with an explicit stabilizer update and spatial-distance analysis. Other protected readout procedures may offer different connectivity or ancillary-qubit requirements; the retained cycle does not establish that this interface is optimal. Its specific contribution is to show how the architecture can be achieved using the matched code geometry and available transversal operations.

\section{Discussion}
Persistent mediation changes how complementary code capabilities can be organized in a fault-tolerant processor. Dimensional code switching accesses the higher-dimensional gate set by changing the encoding of the computational register~\cite{code_switching,code_switching_2,code_switching_3,Beverland_Kubica_Svore_Cost}, while the teleportation-based schemes use an auxiliary encoded block that is eventually read out~\cite{code_switching_with_CNOT,code_switching_with_CNOT_2,code_switching_with_CNOT_3,Lattice_surgery_1}. In our architecture, the 2D data register and 3D mediator remain distinct encoded modules at the boundaries of every logical cycle, with a protected joint subsystem formed only during the cross-dimensional operation. The membrane interface closes this cycle while preserving the protected logical content and spatial distance. The resulting architecture separates the lifetime of an encoded module from the gate capability that it supplies: a higher-dimensional non-Clifford module can remain part of the error-corrected processor and serve successive logical operations.

The construction applies under a definite set of conditions. It requires matched 2D and 3D code geometries, the cross-dimensional interactions used by the interface, continued syndrome extraction, and sufficient reuse to justify maintaining the 3D block between calls. Persistent mediation therefore does not imply a uniform reduction in overhead. Two surgery stages, continuous protection of the mediator, and the associated decoding cost compete with fresh-state preparation, verification, and possible offline parallelism in preparation-based schemes~\cite{Beverland_Kubica_Svore_Cost,magic_cultivation}. The quantitative evidence is also limited to the Clifford surrogate and the noise models and decoder studied here. Exact simulation of the noisy non-Clifford cycle, and a complete circuit-level fault-distance analysis remain to be established. These limitations affect the operating regime and quantitative comparison, while the closed logical cycle and the protected interface construction do not rely on the surrogate.

The smallest matched realization is the seven-qubit Steane code coupled to the fifteen-qubit Reed--Muller code, providing a direct setting in which the closed cycle and mediator recovery can be tested. A minimal validation would verify both the logical action on the 2D register and the return of the 3D mediator over repeated calls, followed by tests with intervening Clifford operations or with the mediator serving different data registers. The next steps are therefore well defined: establish the circuit-level fault distance of the true non-Clifford cycle, determine the resource break-even regime under hardware-specific connectivity and timing constraints, and test composition beyond repeated copies of the same cycle. More broadly, persistent mediation provides a route toward modular fault-tolerant architectures in which distinct encoded modules retain their native gate capabilities and interact through protected interfaces.

\appendix
\section{Physical membrane-surgery construction}

Before introducing the procedure for membrane-like lattice surgery, we must specify the choice of generators for the 3D tetrahedral color code, an essential step in designing the merged lattice. For $d=3$, the lattice contains 18 face operators but only 10 independent $Z$-type stabilizers, leaving eight independent linear dependencies. Similar redundancies occur at larger distances. We choose a boundary-adapted generating set in which every $Z$-type generator is either disjoint from $M_K$, fully contained in $M_K$, or intersects $M_K$ along a single interior membrane edge. The required boundary-adapted choice of $Z$ generators can be made for the standard odd-distance tetrahedral family considered here. In the dual representation, primal $Z$-type face stabilizers correspond to edge operators, and these generators possess local redundancies: at an interior dual vertex, multiplying all incident edge operators of a fixed color pair gives the same vertex stabilizer as multiplying the corresponding incident edges of another color
pair~\cite{Beverland_Kubica_Svore_Cost}. These local relations allow a face generator whose intersection with $M_K$ is an undesired perimeter edge to be replaced by products of face generators that are either disjoint from $M_K$, contained completely in $M_K$, or intersect $M_K$ along an interior membrane edge.
Therefore these three classes span the full $Z$-stabilizer subgroup, and a maximal independent subset can always be chosen as the generator basis used below. We also define interior edges $E_{\rm in}(M)$ as the set of membrane edges incident on exactly two membrane faces, $E_{in}(M) = \Big\{ e \in E(M): \big| \{ f \in M^{(2)} : e \subset \partial f \} \big| = 2 \Big\}$, where $E$ denotes the whole set of edges in the membrane for clarity. We refer to the latter two interface classes as the A-type and B-type generators,
\begin{enumerate}
    \item $A$-type: $A$-type generators of the 2D code are generators whose constituent qubits lie entirely within the membrane. Therefore all generators of the 2D code belong to $A$-type, denoted by $S^P_f$. $A$-type generators of the 3D code are selected generators whose intersection with $M_K$ is exactly a face, denoted by $K^P_f$.
    \item $B$-type: Only the 3D color code involves this type of stabilizer. $B$-type generators are selected face stabilizers whose intersection with $M_K$ is exactly one interior membrane edge $e\in E_{in}(M_K)$. We denote these stabilizers as $K^P_e$.
\end{enumerate}

Next, we introduce the design of the membrane-like surgery protocol. Consider a 2D code $S$ and a 3D code $K$ with respective logical states $\ket{\psi_S}$ and $\ket{\psi_K}$, and corresponding membrane-supported logical Pauli operators $P_L^S$ and $P_L^K$. At the logical level, the merge implements a measurement of $P_L^S\otimes P_L^K$, projecting the two-code logical space with $(I\pm P_L^S\otimes P_L^K)/2$. The physical construction below realizes this operation through a merge-split stabilizer deformation. Using the $P_L^S \otimes P_L^K$ measurement as our primary example, we can describe this process mathematically: lattice surgery between $S$ and $K$ consists of a sequence of topological deformations—mapping the disjoint code spaces $\mathcal{H}_S \otimes \mathcal{H}_K$ to a merged space $\mathcal{H}_{(S \cup K)}$ and back. This effectively applies the projective map $(I \pm P_L^S \otimes P_L^K)/2$ to the joint logical state. We outline this procedure in five distinct steps:

\textbf{Interface-ancilla initialization}. 
For the $X_L\otimes X_L$ membrane surgery, all interface ancillas are initialized in $\ket{0}$. A-type ancillas are placed on membrane faces, while B-type ancillas are placed on the interior membrane edges used by the selected B-type generators. 
\begin{enumerate}
    \item $A$-type: For each membrane face $f$, we place an ancilla $a_f$ at the face center. Each $a_f$ is initialized in $|0\rangle$ and contributes the initialization stabilizer $Z_{a_f}$.
    \item $B$-type: For each selected interior membrane edge $e$, we place an ancilla $b_e$ at the edge midpoint. Each $b_e$ is initialized in $|0\rangle$ and contributes the initialization stabilizer $Z_{b_e}$.
\end{enumerate}
For the selected 2D membrane, only A-type interface checks occur, whereas the 3D membrane contains both A-type and B-type checks. Consequently, the total number of added ancilla qubits corresponds to the sum of the 2-cells and the interior edges of the membrane, yielding $n_I = 3(d - 1)^2/2$. Because the initial logical states reside entirely within the tensor product space of the disjoint codes, appending these ancillas simply expands the overall Hilbert space without altering the encoded quantum information. The joint state can thus be expressed as $\ket{\psi_S} \otimes \ket{\psi_K} \otimes \ket{\psi_A}$, where $\ket{\psi_A}=\ket{0}^{\otimes n_I}$.

\textbf{Modifying the $Z$-type Stabilizers}. 
To merge the two codes, we must first deform their $Z$-type stabilizers into a merged configuration. Similar to standard string-like lattice surgery, our goal is to extend the $Z$-type stabilizers of both membranes so that they incorporate the ancilla qubits, thereby forming new joint stabilizers. In the chosen 2D membrane geometry, the relevant face checks are all $A$-type stabilizers. Each $Z$-type stabilizer expands to include the $A$-ancilla located at its center, yielding the modified stabilizer:
\begin{eqnarray}
\label{Z_stabilizer_enlarge_2D}
S^{'Z}_f = S^Z_f Z_{a_f},
\end{eqnarray}
for every face $f \in M_S$. The corresponding 3D $Z$-type checks are modified in the same way,
\begin{eqnarray}
\label{Z_stabilizer_enlarge_3D_A}
K^{'Z}_f = K^Z_f Z_{a_f},
\end{eqnarray}
for every face $f \in M_K$. Each B-type 3D check is extended to include the corresponding B-type ancilla,
\begin{eqnarray}
\label{Z_stabilizer_enlarge_3D_B}
K^{'Z}_e = K^Z_e Z_{b_e}.
\end{eqnarray}
where $K_e^Z$ denotes the selected B-type face generator associated with the interior edge $e\in M_K$.

\textbf{Creating New $X$-type Stabilizers}. 
After the interface $Z$-type stabilizers are modified, we create new $X$-type stabilizers to close the gap between the 2D and 3D membranes. The construction is most conveniently described from the viewpoint of the 2D membrane vertices. For each vertex $v\in M_S$, let $\pi_0 (v)$ denote the matched vertex on the 3D membrane $M_K$. We introduce one new $X$-type check associated with $v$.

Let $F(v)$ be the set of membrane faces incident on $v$, and let $E(\pi_0 (v))$ be the set of available interface edges incident on $\pi_0 (v)$ that carry $B$-type ancillas. The new $X$-type stabilizer is defined as
\begin{eqnarray}
G_v=X_v X_{\pi_0 (v)}\prod_{f\in F(v)} X_{a_f}\prod_{e\in E(\pi_0(v))} X_{b_e}.
\end{eqnarray}
Here $a_f$ is the $A$-type ancilla associated with the membrane face $f$, and $b_e$ is the $B$-type ancilla associated with the interface edge $e$.

This single rule covers the bulk, boundary, and corner cases. If $v$ is a bulk vertex, all local faces and interface edges around it are present. If $v$ lies on a boundary or a corner, some of these local faces or edges are absent; the products are then simply taken over the available local connections. Thus the corner, boundary, and bulk checks differ only by the number of local ancillas included in the products.

We now verify that the new $X$-type checks commute with the modified $Z$-type checks defined in the previous paragraph. First consider a modified 2D $A$-type check $S_f^{\prime Z}$. This check contains the original 2D face support of $S_f^Z$ together with the ancilla $a_f$. The new check $G_v$ overlaps with the data part of $S_f^{\prime Z}$ exactly when $v\in f$. In the same case, $G_v$ also contains the ancilla $a_f$. Therefore, the two checks overlap on either zero or two qubits and hence commute:
\begin{eqnarray}
[G_v,S_f^{\prime Z}]=0 .
\end{eqnarray}

The same argument applies to the modified 3D $A$-type check $K_f^{\prime Z}$. The matched 3D vertex $\pi_0 (v)$ lies in the corresponding 3D face exactly when $v$ lies in the matched 2D face $f$. Whenever this data-qubit overlap occurs, the new check $G_v$ also contains the same $A$-type ancilla $a_f$. Thus the total overlap parity is again even, 
\begin{eqnarray}
[G_v,K_f^{\prime Z}]=0 .
\end{eqnarray}

Next consider a modified $B$-type check $K_e^{\prime Z}$. The $B$-type check is associated with an interface edge $e$ and includes the ancilla $b_e$. If the matched vertex $\pi_0 (v)$ is incident on $e$, then $G_v$ overlaps with the data part of $K_e^{\prime Z}$ and also contains $b_e$. If $\pi_0 (v)$ is not incident on $e$, there is no such local overlap. Therefore,
\begin{eqnarray}
[G_v,K_e^{\prime Z}]=0.
\end{eqnarray}
All remaining stabilizers are either unchanged stabilizers of the original 2D or 3D color codes, or have no odd overlap with the newly created interface checks. Consequently, the modified $Z$-type checks and the new $X$-type checks form a mutually commuting stabilizer set. Together with the unchanged bulk checks, these operators define the stabilizer-measurement schedule used during the merged stage.

\textbf{Merging the Codes}. 
We next measure the merged-stage check set to implement the stabilizer update. The primary purpose of measuring these modified and newly created stabilizers is to update the stabilizer group, adapting it to the expanded Hilbert space. Notice each $A$-type ancilla $a_f$ appears once for every vertex of the face $f$, and the faces used in the color-code membrane have even weight. Hence all $A$-type ancillas cancel. Each $B$-type ancilla $b_e$ appears twice, once for each endpoint of the edge $e$, and therefore all $B$-type ancillas also cancel. Thus, the product of all new $X$-type checks gives the desired joint logical parity,
\begin{eqnarray}
\prod_{v\in M_S^{(0)}}G_v
=
X(M_S)X(M_K)\equiv \bar X_S \bar X_K.
\end{eqnarray}

Thus, their product yields the joint logical parity $X_L^S \otimes X_L^K$, so the corresponding measurement implements the projector $(I\pm X_L^S X_L^K)/2$ without destructive data readout.

\textbf{Error Correction}. 
Because the stabilizer set changes at the merge and split boundaries, fault tolerance must be analyzed in spacetime. The pre-merge syndrome history together with the initialized interface ancillas defines the initial temporal boundary, while the restored separated-code checks define the terminal boundary after the split. Repeating the merged-stage measurements for $r$ rounds gives the phenomenological spacetime bound $d_{\rm st}\geq\min(d,r)$. We therefore choose $r=d$, the minimum repetition depth for which the temporal direction does not reduce the designed distance.  The split operation is implemented by restoring the original stabilizer measurements of the separated 2D and 3D codes and measuring the interface ancillas in the appropriate basis. In the noiseless setting this is the inverse deformation of the merge. The split forms the terminal boundary of the same repeated-measurement decoding problem and is protected using the same $d$-round repetition depth.

\section{Simulation Details}
We introduce the error models used in the paper, which includes both phenomenological error model and circuit-level error model. For phenomenological tests, a single-qubit depolarizing channel
\begin{equation}
 \mathcal D_p(\rho)=(1-p)\rho+\frac p3\sum_{P\in\{X,Y,Z\}}P\rho P
\end{equation}
acts on data qubits between syndrome rounds, and on participating qubits after transversal gate layers. To characterize measurement errors, each reported stabilizer outcome flips independently with probability $p$. Syndrome extraction itself is otherwise noiseless and instantaneous. 

The circuit-level model involve syndrome extraction circuits. It instead applies single-qubit depolarizing noise of total probabilities $p_i$ and $p_s$ at idle and every single-qubit-gate locations, respectively. It also applies two-qubit depolarizing noise
\begin{equation}
 \mathcal D^{(2)}_{p_t}(\rho)=(1-p_t)\rho+
 \frac{p_t}{15}\sum_{P\in\{I,X,Y,Z\}^{\otimes2}\setminus\{II\}}P\rho P
\end{equation}
after each two-qubit gate. Similar with phenomenological error model, reported measurement bits flip with probability $p_m$. For convenience, the reported circuit-level comparisons set $p_i=p_s=p_t=p_m=p$. 

We use the python pack Stim to constructs the Clifford circuits and detector error models, with Sinter used for sampling\cite{stim}. We use the hypergraph union-find decoder implemented within Hyper-Blossom, corresponding to the cited union-find strategy\cite{hypergraphs_decoder,delfosse2021unionfinddecoderquantumldpc}. The detector model retains hyperedges and flag-derived information even for circuit level.

The single-cycle surrogate starts in $\ket+_S\ket+_K$ and tracks $y_S\oplus m_{zz}\oplus m_{zx}$. A failure is disagreement with its deterministic ideal value. The memory simulations start in logical-$Y$ states using noiseless logical $S$ layers and omit surgery over the matched syndrome duration. Sequential tests initialize once, repeat the surrogate $N$ times, and read out an observable selected by the net $S^N$ action. Interface ancillas are reset between merges. These observables test specified components of the logical output, thus a logical error in any $N$ repeated layers would cause the logical failure.

For the reported circuit-level sampling, stop after at least 100 observed failures or $10^6$ shots. With $k$ failures in $n$ shots and $\hat p=k/n$, the two-sided 95\% Wilson interval has,
\begin{equation}
 c=\frac{\hat p+z^2/(2n)}{1+z^2/n},\qquad
 h=\frac{z\sqrt{\hat p(1-\hat p)/n+z^2/(4n^2)}}{1+z^2/n},\quad z=1.96.
\end{equation}
The interval is $[c-h,c+h]$. The reported upper relative error is $(c+h-\hat p)/\hat p$, approximately $21.6\%$ for points stopped after 100 failures in the low-probability regime.

\section{Membrane-like Merging Details}
\label{app:merge_proof}

In the main text, we explicitly construct the interface checks used during the membrane-like merge, verify their mutual commutation, and show that the product of the new $X$-type checks measures $\bar X_S\bar X_K$. We now characterize the complete post-measurement stabilizer group and the protected logical content resulting from this stabilizer update.

\subsection{Logical operators of merged subsystem}
\label{app:Logical_gauge_degrees}

Let $\mathcal S_S$ and $\mathcal S_K$ denote the stabilizer groups of the two separated codes. The 2D code contains face checks $S_f^X$ and $S_f^Z$, while the 3D code contains cell-type $X$ checks $K_c^X$ and $Z$-type checks denoted generically by $K_\alpha^Z$. The 2D and 3D codes each encode one protected logical qubit, with logical operators $\bar X_S,\bar Z_S$ and $\bar X_K,\bar Z_K$.

Before the merge, the two separated code blocks have protected logical Pauli group generated by
\begin{equation}
\bar X_S,\ \bar Z_S,\ \bar X_K,\ \bar Z_K.
\end{equation}
The membrane-like merge measures the joint logical parity $\bar X_S\bar X_K$. Equivalently, in the merged code space, $\bar X_S\bar X_K$ is promoted to a stabilizer of the protected subsystem. Therefore,
\begin{equation}
[\bar X_S]=[\bar X_K],
\end{equation}
where $[\cdot]$ denotes equivalence modulo the complete stabilizer group of the merged state, whose interface contribution is characterized in the next subsection.

Therefore, any protected logical class inherited from the two original blocks can be represented as
\begin{equation}
P_L
=
\bar X_S^{a}
\bar Z_S^{b}
\bar X_K^{c}
\bar Z_K^{d},
\qquad
a,b,c,d\in\{0,1\}.
\end{equation}
For $P_L$ to commute with the newly measured joint parity $\bar X_S\bar X_K$, the number of $Z$-type logical operators that anticommute with $\bar X_S\bar X_K$ must be even. Hence
\begin{equation}
P_L
=
\bar X_S^{a}
\bar X_K^{c}
(\bar Z_S\bar Z_K)^b.
\end{equation}
Using the measured relation $[\bar X_S]=[\bar X_K]$, this becomes$[P_L]=[\bar X_S]^{a+c}[\bar Z_S\bar Z_K]^b$. Therefore the protected logical group of the merged subsystem is generated by
\begin{equation}
\bar X_M=[\bar X_S]=[\bar X_K],
\qquad
\bar Z_M=\bar Z_S\bar Z_K.
\end{equation}
These two operators anticommute because $\bar X_S$ anticommute with $\bar Z_S$, and commutes with $\bar Z_K$. Hence they form one protected logical qubit. Possible logical operators supported entirely on the interface ancillas are analyzed explicitly in the following subsection. 

Although the joint parity is obtained from the product of all newly introduced $X$-type interface checks, the measurement procedure records the outcomes of the individual checks $K_v^{\prime X}$. We therefore verify that these additional outcomes do not reveal the individual protected logical information of either code block. Consider an arbitrary parity of the measured interface outcomes, associated with the operator \begin{eqnarray} 
P_A = \prod_{v\in A}K_v^{\prime X}, \qquad A\subseteq M_S^{(0)} . 
\end{eqnarray} 
From the definition of $K_v^{\prime X}$, the restriction of $P_A$ to the two data membranes has matched $X$ support; the remaining support acts only on interface ancillas: 
\begin{eqnarray}
P_A = X(A)_S X(\widetilde A)_K G_A , 
\end{eqnarray} 
where $X(A)_S=\prod_{v\in A}X_v$, $X(\widetilde A)_K=\prod_{v\in A}X_{\pi_0 (v)}$, and $G_A$ contains only interface-ancilla operators. The matched sets $A$ and $\widetilde A$ contain the same number of vertices. Using the membrane representatives of the logical $Z$ operators established in the main text, the commutation parity of $P_A$ with the two logical $Z$ operators is therefore identical: 
\begin{eqnarray}
P_A\bar Z_S &=& (-1)^{|A|} \bar Z_S P_A , \\ P_A\bar Z_K &=& (-1)^{|A|} \bar Z_K P_A . 
\end{eqnarray}
The interface factor $G_A$ does not affect these relations because it has support only on the added interface ancillas and thus does not overlap with membranes. Now consider those products $P_A$ that, after removing stabilizer and interface-local contributions, have a well-defined action on the protected logical subsystem. Since the measured interface operators are $X$-type, their possible protected logical classes in the two original code blocks are identity, $\bar X_S, \bar X_K$, and their product $\bar X_S \bar X_K$.
These classes are distinguished by their commutation parities with $\bar Z_S$ and $\bar Z_K$. The classes $\bar X_S$ and $\bar X_K$ have different commutation parities with the two logical $Z$ operators, whereas every $P_A$ constructed above has identical parity with them. Since $\bar X_S$ and $\bar X_K$ have different commutation signatures with $\bar Z_S$ and $\bar Z_K$, neither class can be represented by any $P_A$. Consequently, 
\begin{eqnarray}
P_A \not\equiv \bar X_S, \qquad P_A \not\equiv \bar X_K \end{eqnarray}
on the protected logical subsystem. The only possible protected logical classes accessible from products of the interface measurement outcomes are therefore the trivial class and the joint class $\bar X_S\bar X_K$. As shown in the main text, the product of all new $X$-type checks realizes this nontrivial case, 
\begin{eqnarray}
\prod_{v\in M_S^{(0)}}K_v^{\prime X} \equiv \bar X_S\bar X_K . \end{eqnarray}
Hence the individual interface outcomes may contain local syndrome or interface information, but they do not separately reveal the logical state of either the 2D or the 3D code block. Therefore at the protected logical level, the only nontrivial information extracted by the merge is the intended joint parity $\bar X_S\bar X_K$.

\subsection{Stabilizer update and interface-local constraints}
\label{app:stabilizer_update}

The previous subsection identifies the protected logical information extracted by the membrane-like merge. We now clarify the stabilizer-rank change during the merge and show explicitly that the added interface ancillas do not introduce additional unknown logical degrees of freedom.

Let $n_S$ and $n_K$ denote the numbers of physical qubits in the 2D and 3D code blocks, respectively, and let $n_I$ denote the number of added interface ancillas. During the merge, the total number of physical qubits is $N=n_S+n_K+n_I$.

Both original code blocks encode one logical qubit. Their stabilizer ranks are therefore $n_S-1$ and $n_K-1$. In addition, every interface ancilla is initialized in $\ket{0}$ and contributes one independent single-qubit $Z$ stabilizer. Before the merge measurements are performed, the complete stabilizer group is
\begin{eqnarray}
\mathcal S_{\rm pre}
=
\left\langle
\mathcal S_S,
\mathcal S_K,
\left\{
Z_{\alpha}:\alpha\in\mathcal I
\right\}
\right\rangle ,
\label{eq:pre_merge_group}
\end{eqnarray}
where $\mathcal I$ denotes the interface-ancilla set. Its rank is
${\rm rank}(\mathcal S_{\rm pre})=(n_S-1)+(n_K-1)+n_I=N-2$. Hence the pre-merge system encodes two logical qubits.
The interface ancillas therefore introduce no additional encoded qubits before the merge. Each additional physical qubit is fixed by an independent initialization stabilizer.

The modified $Z$-type checks introduced in the main text are already elements of $\mathcal S_{\rm pre}$. For example, $S_f^{\prime Z}=S_f^Z Z_{a_f}$ and $K_e^{\prime Z}= K_e^Z Z_{b_e}$ are products of stabilizers present before the merge. Their role is not to increase the stabilizer rank, but to provide combinations of the pre-merge constraints that commute with the newly measured $X$-type interface checks. The individual factors, such as $S_f^Z$ and $Z_{a_f}$, need not separately remain stabilizers after the new $X$ measurements are performed.
Let
\begin{eqnarray}
\mathcal X_{\rm merge}
=
\left\langle
K_v^{\prime X}:
v\in M_S^{(0)}
\right\rangle
\label{eq:X_merge_group}
\end{eqnarray}
be the group generated by the new $X$-type interface checks. Ignoring the signs determined by their measurement outcomes, the post-measurement stabilizer group is obtained by the standard stabilizer update,
\begin{eqnarray}
\mathcal S_{\rm post}
=
\left\langle
\mathcal X_{\rm merge},
\mathcal S_{\rm pre}\cap
C(\mathcal X_{\rm merge})
\right\rangle ,
\label{eq:post_merge_group}
\end{eqnarray}
where $C(\mathcal X_{\rm merge})$ denotes the Pauli centralizer of the measured group.

For a commuting family of Pauli measurements, the increase of the stabilizer rank is equal to the number of independent measured operators that commute with the pre-measurement stabilizer group but are not already stabilizers. In the present notation,
\begin{eqnarray}
{\rm rank}(\mathcal S_{\rm post})
-
{\rm rank}(\mathcal S_{\rm pre})
=
\dim_{\mathbb F_2}
\frac{
\mathcal X_{\rm merge}
\cap
C(\mathcal S_{\rm pre})
}{
\mathcal X_{\rm merge}
\cap
\mathcal S_{\rm pre}
}.
\label{eq:rank_update_general}
\end{eqnarray}
Measurements outside $C(\mathcal S_{\rm pre})$ replace anticommuting pre-existing stabilizer generators and therefore do not change the stabilizer rank.

The quotient in Eq. (\ref{eq:rank_update_general}) has already been characterized by the logical-information analysis in the previous subsection. Any product of the new $X$-type checks that has a well-defined action on the protected logical sector can represent only the trivial logical class or the joint class $\bar X_S\bar X_K$. Moreover, the product of all new $X$-type checks realizes the nontrivial class,
\begin{eqnarray}
\prod_{v\in M_S^{(0)}}
K_v^{\prime X}
\equiv
\bar X_S\bar X_K .
\label{eq:joint_logical_constraint}
\end{eqnarray}
Therefore,
\begin{eqnarray}
\dim_{\mathbb F_2}
\frac{
\mathcal X_{\rm merge}
\cap
C(\mathcal S_{\rm pre})
}{
\mathcal X_{\rm merge}
\cap
\mathcal S_{\rm pre}
}
=
1 .
\label{eq:logical_measurement_dimension}
\end{eqnarray}
It follows that ${\rm rank}(\mathcal S_{\rm post})={\rm rank}(\mathcal S_{\rm pre})+1$ and consequently $k_{\rm post}=N-{\rm rank}(\mathcal S_{\rm post})=1$. This gives a direct stabilizer interpretation of the merge. The local interface measurements replace pre-existing stabilizer constraints without changing the encoded dimension, while the single independent joint-logical measurement $\bar X_S\bar X_K$ adds one new stabilizer constraint. The protected logical content therefore changes from $k_{\rm pre}=2$ to $k_{\rm post}=1$. No additional unknown logical degree of freedom is generated by the interface.

For completeness, we also show that no hidden logical operator can be supported entirely on the interface. Introduce the binary incidence matrix $\Gamma_{\rm int}$ whose rows are indexed by the new checks $K_v^{\prime X}$ and whose columns are indexed by the interface ancillas. For an $A$-type ancilla $a_f$ and a $B$-type ancilla $b_e$,
\begin{eqnarray}
\Gamma_{\rm int}[v,a_f]
&=&
[v\in f],
\\
\Gamma_{\rm int}[v,b_e]
&=&
[\pi_0(v)\in e],
\label{eq:Gamma_interface}
\end{eqnarray}
where $[\cdot]$ denotes the binary indicator function.

A product of initial interface-$Z$ stabilizers can be written as
\begin{eqnarray}
Z(\mathbf z_I)
=
\prod_{\alpha\in\mathcal I}
Z_{\alpha}^{(\mathbf z_I)_{\alpha}} .
\end{eqnarray}
Its commutation relation with $K_v^{\prime X}$ is
\begin{eqnarray}
K_v^{\prime X}Z(\mathbf z_I)
=
(-1)^{(\Gamma_{\rm int}\mathbf z_I)_v}
Z(\mathbf z_I)K_v^{\prime X}.
\end{eqnarray}
Hence the purely interface-supported initialization constraints that survive unchanged are
\begin{eqnarray}
\mathcal S_I^{\rm surv}
=
\left\{
Z(\mathbf z_I):
\Gamma_{\rm int}\mathbf z_I=0
\pmod 2
\right\}.
\label{eq:S_interface_surv}
\end{eqnarray}
These operators belong to $\mathcal S_{\rm pre}$ and commute with the complete measurement group, and therefore belong to the post-measurement stabilizer group $\mathcal S_{\rm post}$.

Finally, we exclude a nontrivial logical Pauli supported entirely on the interface. Consider an interface-supported Pauli operator$P_I=X(\mathbf x_I)Z(\mathbf z_I)$ 
that commutes with $\mathcal S_{\rm post}$. Each interface ancilla appears as the unique interface-ancilla factor in at least one modified $Z$-type stabilizer. Therefore, any nonzero $X$ component on an interface ancilla anticommutes with one such stabilizer, implying $\mathbf x_I=0$. Thus $P_I=Z(\mathbf z_I)$.

Commutation with all new $X$-type checks requires
$\Gamma_{\rm int}\mathbf z_I=0\pmod 2$. Eq. (\ref{eq:S_interface_surv}) then implies
\begin{eqnarray}
P_I
\in
\mathcal S_I^{\rm surv}
\subseteq
\mathcal S_{\rm post}.
\end{eqnarray}
Therefore, $P_I\in\mathcal S_{\rm post}$. There is no nontrivial logical Pauli operator supported entirely on the interface subsystem.

\subsection{Protected distance of the merged code}
\label{app:protected_distance}

We next show that the membrane-like interface does not reduce the spatial distance of the protected logical qubit. The preceding subsection excludes a logical operator supported entirely on the interface. It remains to exclude a lower-weight logical representative whose support is distributed across the 2D code, the 3D code, and the interface ancillas.

Because the post-merge code is CSS, it is sufficient to bound the $X$- and $Z$-type logical sectors separately. We first consider the $Z$-type logical sector of the merged code.
The protected logical $Z$ class after the merge is
\begin{eqnarray}
\bar Z_M
=
\bar Z_S\bar Z_K .
\label{eq:Z_M}
\end{eqnarray}
Let $L_Z$ be an arbitrary $Z$-type representative of this logical class. We define two restriction maps, $\mathcal R_Z^S$ and $\mathcal R_Z^K$, which retain the support of a $Z$-type Pauli operator on the original 2D and 3D data blocks, respectively, while removing the support on the other code block and on the interface ancillas.

The $Z$-type stabilizers of the post-merge code are inherited from the pre-merge $Z$-stabilizer group through combinations that commute with the new $X$-type measurements. Therefore, the restriction of every post-merge $Z$ stabilizer to either original code block belongs to the corresponding original $Z$-stabilizer group or is the identity. Consequently,
\begin{eqnarray}
\mathcal R_Z^S(L_Z)
&\equiv&
\bar Z_S
\qquad
{\rm mod}\ \mathcal S_S^Z,
\\ 
\mathcal R_Z^K(L_Z)
&\equiv&
\bar Z_K
\qquad
{\rm mod}\ \mathcal S_K^Z .
\label{eq:Z_logical_restrictions}
\end{eqnarray}
Thus both restrictions are nontrivial logical $Z$ operators of their respective original codes, and
\begin{eqnarray}
{\rm wt}\left(\mathcal R_Z^S(L_Z)\right)
&\geq&
d_S^Z,
\\
{\rm wt}\left(\mathcal R_Z^K(L_Z)\right)
&\geq&
d_K^Z .
\end{eqnarray}

Since the two original code blocks have disjoint physical support, their contributions to the weight are additive. Any additional support on the interface ancillas can only increase the total weight. Hence
\begin{eqnarray}
{\rm wt}(L_Z)
&\geq&
{\rm wt}\left(\mathcal R_Z^S(L_Z)\right)
+
{\rm wt}\left(\mathcal R_Z^K(L_Z)\right)
\nonumber\\
&\geq&
d_S^Z+d_K^Z,
\label{eq:Z_distance_lower}
\end{eqnarray}
which leads to $d_M^Z\geq d_S^Z+d_K^Z$.

Next we consider the $X$-type logical sector of the merged code.
The protected logical $X$ class satisfies
\begin{eqnarray}
\bar X_M
=
[\bar X_S]
=
[\bar X_K].
\label{eq:X_M}
\end{eqnarray}
To exclude a mixed-region shortcut, we define a reduction map from an $X$-type Pauli operator on the merged system to an $X$-type Pauli operator on the original 2D code.

Let $\mathbf x_S$ and $\mathbf x_K$ denote the binary $X$ supports on the 2D and 3D data blocks. For every matched pair of membrane vertices $v$ and $\pi_0(v)$, define
\begin{eqnarray}
\left[
\mathcal R_X(\mathbf x)
\right]_v
=
(\mathbf x_S)_v
+
(\mathbf x_K)_{\pi_0(v)}
\pmod 2 .
\label{eq:RX_definition}
\end{eqnarray}
All support on the interface ancillas and on 3D bulk vertices away from the matched membrane is discarded.

The reduction map is compatible with the post-merge $X$-type stabilizer group. An original 2D $X$ stabilizer is mapped to the same 2D stabilizer. For a 3D $X$-type cell stabilizer, only its intersection with the matched boundary membrane contributes. By the membrane-intersection property established in the cross-dimensional construction, the pullback of this intersection belongs to the 2D $X$-stabilizer group. Hence
\begin{eqnarray}
\mathcal R_X(K_c^X)
\in
\mathcal S_S^X .
\label{eq:RX_3D_stabilizer}
\end{eqnarray}
For each new interface check $K_v^{\prime X}$, the 2D vertex $v$ and its matched 3D vertex $\pi_0(v)$ both appear once. Their contributions cancel under the binary reduction, while the interface-ancilla factors are discarded. Thus every post-merge $X$-type stabilizer is mapped to a 2D $X$ stabilizer or to the identity.

Let $L_X$ be an arbitrary representative of the protected logical $X$ class. Since $L_X$ differs from $\bar X_S$ only by post-merge $X$ stabilizers,
\begin{eqnarray}
\mathcal R_X(L_X)
\equiv
\bar X_S
\qquad
{\rm mod}\ \mathcal S_S^X .
\label{eq:RX_logical}
\end{eqnarray}
Therefore, $\mathcal R_X(L_X)$ is a nontrivial logical $X$ operator of the original 2D color code.

The map $\mathcal R_X$ cannot increase the Pauli weight. A nonzero reduced support on a 2D vertex requires an odd occupation of the corresponding matched pair $\{v,\pi_0(v)\}$, and different output vertices correspond to disjoint matched pairs. All other support is discarded. Consequently,
\begin{eqnarray}
{\rm wt}\left(\mathcal R_X(L_X)\right)
\leq
{\rm wt}(L_X).
\label{eq:RX_weight}
\end{eqnarray}
It follows that
\begin{eqnarray}
{\rm wt}(L_X)
\geq
d_S^X .
\label{eq:X_distance_lower}
\end{eqnarray}

Therefore, a minimum-weight representative of the original 2D logical operator $\bar X_S$ remains a valid post-merge logical operator. Every post-merge $Z$ stabilizer restricts to a 2D $Z$ stabilizer or identity; therefore $\bar X_S$ commutes with all post-merge $Z$ stabilizers. Moreover, since $\bar Z_M=\bar Z_S\bar Z_K$, $\bar X_S$ anticommutes with $\bar Z_M$ and is therefore not a stabilizer. Hence
\begin{eqnarray}
d_M^X
=
d_S^X .
\label{eq:X_distance_exact}
\end{eqnarray}

Because the merged code is CSS, the spatial distance of the protected post-merge logical qubit is
\begin{eqnarray}
d_M
=
\min\left\{
d_M^X,d_M^Z
\right\}
=
d .
\label{eq:merged_distance_final}
\end{eqnarray}

\subsection{Repeated syndrome measurements and spacetime protection} 

The preceding analysis establishes the spatial distance $d_M$ of the protected logical qubit during the merged stage. We now consider measurement errors by repeating the merged stabilizer measurements for $r$ rounds. In the phenomenological noise model, data faults may occur between consecutive syndrome-extraction rounds, while measurement faults may flip individual stabilizer outcomes. For each stabilizer check, a detector is defined from the change in its measured outcome between adjacent rounds, together with the appropriate initial- and final-time boundary conditions. Data faults therefore generate spatial detector patterns, whereas measurement faults generate time-like detector events. 

The first merge round has a well-defined temporal boundary. Before the merge, the 2D and 3D code stabilizers and the $\ket{0}$ initialization of the interface ancillas provide the reference constraints from which the first merged-check outcomes are inferred. The split operation similarly provides the terminal temporal boundary by restoring the separated-code stabilizer structure. A logical failure of the repeated joint-parity measurement must correspond to a detector-free nontrivial fault configuration connecting the relevant spacetime boundaries. A purely spatial configuration must implement a nontrivial protected logical operator of the merged code and therefore has weight at least $d_M$. A purely time-like sequence capable of changing the inferred parity must extend across the repeated-measurement window and therefore has weight at least $r$. Mixed space-time configurations can contain both data and measurement faults, but they cannot terminate freely inside the merged spacetime region. Their endpoints must either be detected in the bulk or terminate on one of the explicitly defined temporal or logical boundaries. 

In particular, the merge and split interfaces do not provide additional unconstrained endpoints, because the first-round and final-round detector conditions are fixed by the pre-merge initialization and the restored separated-code stabilizers. A nontrivial mixed chain must therefore either project to a nontrivial spatial logical chain of length at least $d_M$ or span the complete temporal interval of $r$ rounds. The shortest nontrivial spacetime fault configuration is therefore bounded by 
\begin{eqnarray}
d_{\rm st} \geq \min\left(d_M,r\right). \label{eq:spacetime_distance} 
\end{eqnarray} 
Choosing $r=d_M$ is the minimum repetition depth for which the temporal direction does not reduce the designed distance of the merged parity measurement. 

Thus, in the phenomenological model used in this work, fewer than $d_M$ faults cannot silently change the inferred value of the joint logical parity $\bar X_S\bar X_K$ without producing a nontrivial detector syndrome. The split is treated as the terminal boundary of the same spacetime decoding problem. In the ideal code-space picture it reverses the stabilizer deformation introduced by the merge, while in the noisy repeated-measurement picture its final measurements close the detector history and restore the stabilizer structure of the separated 2D and 3D code blocks.

\subsection{Circuit-level fault-tolerant implementation} \label{app:circuit_level_ft} 

The preceding spacetime argument treats stabilizer readout at the phenomenological level, where a stabilizer measurement is regarded as an operation whose outcome may be flipped. At circuit-level, syndrome extraction is implemented using noisy state preparation, entangling gates, measurements, and idle operations. A fault occurring on a syndrome ancilla may propagate through subsequent entangling gates and generate a correlated multi-qubit data error, commonly referred to as a hook error. 

For the circuit-level implementation considered in this work, the high-weight stabilizer measurements are realized using flag-based fault-tolerant syndrome-extraction circuits of the type developed in Refs.~\cite{Chao_Reichardt_Two_Extra_Qubits,Chamberland_Beverland_Flag_Arbitrary_Distance,Chao_Reichardt_Flag_Any_Stabilizer_Code,Prabhu_Reichardt_Fewer_Qubits}. The role of the flag ancillas is to identify fault patterns that can otherwise propagate into data errors whose weight exceeds that expected from the number of underlying faults. For a code designed to tolerate $t$ faults, the stabilizer-measurement gadget is chosen so that any fault set of size at most $t$ either produces a data error within the corresponding correctable set or generates a nontrivial flag pattern that restricts the possible correlated errors. 

The flag outcomes are retained as part of the syndrome information and supplied to the decoder; flagged events are not post-selected away. We do not derive the general hook-error analysis of these flag constructions here. Instead, we use the established flag-fault-tolerance conditions for the individual high-weight Pauli measurements and apply the resulting flag-conditioned syndrome information to the merged-code decoding problem. This ensures that the explicit syndrome-extraction gadgets used in our implementation do not produce unflagged high-weight hook errors below the designed fault-tolerance order. The existence of fault-tolerant flag gadgets for the individual stabilizer measurements should be distinguished from a separate analytic proof of the complete circuit-level fault distance of the entire time-dependent merge--split protocol. In this work, the former provides the circuit-level realization of the high-weight checks, while the complete gadget is evaluated numerically using the circuit-level noise model. The corresponding detector graph includes the syndrome and flag outcomes generated by the explicit extraction circuits.

\section{Logical-State Evolution of the $T$-Gate Gadget}

In this appendix, we track the ideal logical state through each step of the $T$-gate gadget, where all intermediate states are written without normalization factors. The initial joint state in the logical code space is:
\begin{eqnarray}
\label{psi_-1}
\ket{\psi_0} = \left( \alpha\ket{0}_S + \beta\ket{1}_S\right) \otimes \ket{+}_K.
\end{eqnarray}
Applying the $CZ$ gate. The $CZ$ gate entangles the two codes by flipping the sign of the $\ket{11}$ term.
\begin{eqnarray}
\label{psi_1}
\ket{\psi_1}=  \alpha\ket{00} + \alpha\ket{01} + \beta\ket{10} - \beta\ket{11}.
\end{eqnarray}
Measure $Z_L\otimes Z_L$ by string-like lattice surgery, where the outcome is denoted by $m_{zz}\in \{0, 1\}$.
\begin{eqnarray}
\label{psi_2}
\ket{\psi_2} = \alpha \ket{0, m_{zz}}- (-1)^{m_{zz}} \beta \ket{1, 1 \oplus m_{zz}}.
\end{eqnarray}
Apply the $T$ gate on the 3D code. The $T$ gate injects a phase factor $e^{i\pi/4}$ when the 3D code is in the state $\ket{1}_L$.
\begin{align}
\label{psi_3}
\ket{\psi_3}=& \alpha e^{i\frac{\pi}{4} m_{zz}} \ket{0, m_{zz}}  \nonumber \\&- (-1)^{m_{zz}} \beta e^{i\frac{\pi}{4} (1 \oplus m_{zz})} \ket{1, 1 \oplus m_{zz}}.
\end{align}
Measure $Z_L\otimes X_L$ by membrane-like lattice surgery, where the outcome is denoted by $m_{zx}\in \{0, 1\}$.
\begin{align}
\label{psi_4}
\ket{\psi_4} = \alpha'\ket{0}_{S} \otimes Z^{m_{zx}}\ket+_{K}  + \beta' \ket{1}_{S} \otimes Z^{m_{zx}}\ket{-}_{K},
\end{align}
where $\alpha'=\alpha (-1)^{m_{zz} \cdot m_{zx}} e^{i\frac{\pi}{4} m_{zz}}$ and $\beta'=\beta (-1)^{(1 \oplus m_{zz}) \cdot m_{zx}} e^{i\frac{\pi}{4} (1 \oplus m_{zz})}$

Apply the feed-forward corrections. We write the relative phase factor as 
\begin{eqnarray}
\label{psi_2}
\frac{\alpha'}{\beta'}/\frac{\alpha}{\beta} = (-1)^{m_{zx}} e^{i\frac{\pi}{4}(-1)^{m_{zz}}}.
\end{eqnarray}

We apply the outcome-dependent Clifford corrections listed in Table \ref{feed-forward}, which yields
\begin{equation}
\ket{\psi_5} =\alpha \ket{0}_S\ket{+}_K+\beta e^{i\pi/4}\ket{1}_S\ket{-}_K.
\end{equation}

Apply the final $CZ$ gate. The final $CZ$ disentangles the two code blocks, yielding the logical $T$ operation on the 2D code while returning the 3D mediator to $\ket{+}_K$.
\begin{eqnarray}
\label{psi_6}
\ket{\psi_6} = \left( \alpha\ket{0}_{S} + \beta e^{i\pi/4}\ket{1}_{S} \right) \otimes \ket{+}_{K}.
\end{eqnarray}
Thus, after the prescribed feed-forward correction, the ideal closed cycle acts as $T_S\otimes I_K$ on states of the form $\ket{\psi}_S\otimes\ket{+}_K$.

\begin{table}[htbp]
\centering
\caption{Logical feed-forward corrections.}
\label{feed-forward}
\begin{tabular}{@{}cccc@{}}
\hline
\(a=m_{zz}\) & \(b=m_{zx}\) & Correction on 2D block & Correction on 3D block\\
\hline
0 & 0 & \(I\) & \(I\)\\
0 & 1 & \(Z\) & \(Z\)\\
1 & 0 & \(S\) & \(I\)\\
1 & 1 & \(S^\dagger\) & \(Z\)\\
\hline
\end{tabular}
\end{table}

\section{Stabilizer Lists and Flag-Circuit Sequences}

\begin{table*}[htbp]
\centering
\begin{tabular}{| p{1.5cm} | p{6cm} | p{1.5cm} | p{6cm} |}
\hline
\multicolumn{4}{|c|}{$Z$-type Stabilizers} \\
\hline
$K^Z_{1}$ & $Z_{0} Z_{13} Z_{35} Z_{56}$ & $K^Z_{2}$ & $Z_{6} Z_{13} Z_{25} Z_{56}$ \\
\hline
$K^Z_{3}$ & $Z_{4} Z_{6} Z_{21} Z_{25} Z_{33} Z_{55}$ & $K^Z_{4}$ & $Z_{33} Z_{41} Z_{55} Z_{58}$ \\
\hline
$K^Z_{5}$ & $Z_{4} Z_{13} Z_{14} Z_{25} Z_{31} Z_{35}$ & $K^Z_{6}$ & $Z_{7} Z_{14} Z_{31} Z_{54}$ \\
\hline
$K^Z_{7}$ & $Z_{5} Z_{7} Z_{40} Z_{54}$ & $K^Z_{8}$ & $Z_{4} Z_{7} Z_{19} Z_{21} Z_{31} Z_{40}$ \\
\hline
$K^Z_{9}$ & $Z_{19} Z_{21} Z_{55} Z_{58}$ & $K^Z_{10}$ & $Z_{38} Z_{55} Z_{58} Z_{61}$ \\
\hline
$K^Z_{11}$ & $Z_{22} Z_{33} Z_{38} Z_{55}$ & $K^Z_{12}$ & $Z_{18} Z_{21} Z_{32} Z_{38} Z_{47} Z_{55}$ \\
\hline
$K^Z_{13}$ & $Z_{19} Z_{21} Z_{32} Z_{43}$ & $K^Z_{14}$ & $Z_{1} Z_{6} Z_{25} Z_{59}$ \\
\hline
$K^Z_{15}$ & $Z_{4} Z_{21} Z_{24} Z_{32}$ & $K^Z_{16}$ & $Z_{1} Z_{4} Z_{24} Z_{25} Z_{44} Z_{60}$ \\
\hline
$K^Z_{17}$ & $Z_{1} Z_{13} Z_{25} Z_{27}$ & $K^Z_{18}$ & $Z_{4} Z_{17} Z_{24} Z_{31}$ \\
\hline
$K^Z_{19}$ & $Z_{7} Z_{37} Z_{40} Z_{46}$ & $K^Z_{20}$ & $Z_{13} Z_{20} Z_{27} Z_{56}$ \\
\hline
$K^Z_{21}$ & $Z_{7} Z_{17} Z_{31} Z_{37}$ & $K^Z_{22}$ & $Z_{13} Z_{27} Z_{35} Z_{64}$ \\
\hline
$K^Z_{23}$ & $Z_{14} Z_{17} Z_{31} Z_{34}$ & $K^Z_{24}$ & $Z_{7} Z_{10} Z_{37} Z_{54}$ \\
\hline
$K^Z_{25}$ & $Z_{2} Z_{12} Z_{26} Z_{63}$ & $K^Z_{26}$ & $Z_{3} Z_{18} Z_{39} Z_{47}$ \\
\hline
$K^Z_{27}$ & $Z_{8} Z_{23} Z_{45} Z_{50}$ & $K^Z_{28}$ & $Z_{9} Z_{15} Z_{49} Z_{51}$ \\
\hline
$K^Z_{29}$ & $Z_{11} Z_{12} Z_{52} Z_{60}$ & $K^Z_{30}$ & $Z_{2} Z_{17} Z_{24} Z_{34} Z_{44} Z_{63}$ \\
\hline
$K^Z_{31}$ & $Z_{9} Z_{23} Z_{45} Z_{49}$ & $K^Z_{32}$ & $Z_{16} Z_{18} Z_{39} Z_{57}$ \\
\hline
$K^Z_{33}$ & $Z_{18} Z_{44} Z_{47} Z_{52} Z_{57} Z_{60}$ & $K^Z_{34}$ & $Z_{2} Z_{23} Z_{50} Z_{63}$ \\
\hline
$K^Z_{35}$ & $Z_{20} Z_{27} Z_{30} Z_{64}$ & $K^Z_{36}$ & $Z_{12} Z_{44} Z_{60} Z_{63}$ \\
\hline
$K^Z_{37}$ & $Z_{9} Z_{15} Z_{28} Z_{42}$ & $K^Z_{38}$ & $Z_{11} Z_{12} Z_{26} Z_{29}$ \\
\hline
$K^Z_{39}$ & $Z_{24} Z_{32} Z_{44} Z_{47}$ & $K^Z_{40}$ & $Z_{22} Z_{36} Z_{38} Z_{61}$ \\
\hline
$K^Z_{41}$ & $Z_{10} Z_{37} Z_{46} Z_{53}$ & $K^Z_{42}$ & $Z_{3} Z_{32} Z_{43} Z_{47}$ \\
\hline
$K^Z_{43}$ & $Z_{18} Z_{22} Z_{38} Z_{57}$ & $K^Z_{44}$ & $Z_{9} Z_{42} Z_{48} Z_{49}$ \\
\hline
$K^Z_{45}$ & $Z_{3} Z_{23} Z_{44} Z_{45} Z_{47} Z_{63}$ & $K^Z_{46}$ & $Z_{48} Z_{49} Z_{51} Z_{62}$ \\
\hline
$K^Z_{47}$ & $Z_{23} Z_{49} Z_{50} Z_{51}$ & $K^Z_{48}$ & $Z_{11} Z_{12} Z_{23} Z_{48} Z_{49} Z_{63}$ \\
\hline
\multicolumn{4}{|c|}{$X$-type Stabilizers} \\
\hline
$K^X_{1}$ & \multicolumn{3}{l|}{$X_{0} X_{13} X_{20} X_{27} X_{30} X_{35} X_{56} X_{64}$}\\ 
\hline
$K^X_{2}$ & \multicolumn{3}{|l|}{$X_{1} X_{6} X_{13} X_{20} X_{25} X_{27} X_{56} X_{59}$} \\
\hline
$K^X_{3}$ & \multicolumn{3}{|l|}{$X_{1} X_{4} X_{6} X_{18} X_{21} X_{22} X_{24} X_{25} X_{32} X_{33} X_{38} X_{44} X_{47} X_{52} X_{55} X_{57} X_{59} X_{60}$} \\ 
\hline
$K^X_{4}$ & \multicolumn{3}{|l|}{$X_{22} X_{33} X_{36} X_{38} X_{41} X_{55} X_{58} X_{61}$} \\
\hline
$K^X_{5}$ & \multicolumn{3}{|l|}{$X_{1} X_{2} X_{4} X_{12} X_{13} X_{14} X_{17} X_{24} X_{25} X_{26} X_{27} X_{31} X_{34} X_{35} X_{44} X_{60} X_{63} X_{64}$} \\ 
\hline
$K^X_{6}$ & \multicolumn{3}{|l|}{$X_{7} X_{10} X_{14} X_{17} X_{31} X_{34} X_{37} X_{54}$} \\
\hline
$K^X_{7}$ & \multicolumn{3}{|l|}{$X_{5} X_{7} X_{10} X_{37} X_{40} X_{46} X_{53} X_{54}$} \\ 
\hline
$K^X_{8}$ & \multicolumn{3}{|l|}{$X_{4} X_{7} X_{17} X_{19} X_{21} X_{24} X_{31} X_{32} X_{37} X_{40} X_{43} X_{46}$} \\
\hline
$K^X_{9}$ & \multicolumn{3}{|l|}{$X_{3} X_{18} X_{19} X_{21} X_{32} X_{38} X_{39} X_{43} X_{47} X_{55} X_{58} X_{61}$} \\
\hline
$K^X_{10}$ & \multicolumn{3}{|l|}{$X_{1} X_{11} X_{12} X_{20} X_{26} X_{27} X_{29} X_{30} X_{52} X_{59} X_{60} X_{64}$} \\
\hline
$K^X_{11}$ & \multicolumn{3}{|l|}{$X_{2} X_{3} X_{8} X_{10} X_{17} X_{23} X_{24} X_{32} X_{34} X_{37} X_{43} X_{44} X_{45} X_{46} X_{47} X_{50} X_{53} X_{63}$} \\ 
\hline
$K^X_{12}$ & \multicolumn{3}{|l|}{$X_{16} X_{18} X_{22} X_{36} X_{38} X_{39} X_{57} X_{61}$} \\
\hline
$K^X_{13}$ & \multicolumn{3}{|l|}{$X_{2} X_{11} X_{12} X_{23} X_{26} X_{29} X_{48} X_{49} X_{50} X_{51} X_{62} X_{63}$} \\ 
\hline
$K^X_{14}$ & \multicolumn{3}{|l|}{$X_{8} X_{9} X_{15} X_{23} X_{45} X_{49} X_{50} X_{51}$} \\
\hline
$K^X_{15}$ & \multicolumn{3}{|l|}{$X_{3} X_{9} X_{11} X_{12} X_{16} X_{18} X_{23} X_{39} X_{42} X_{44} X_{45} X_{47} X_{48} X_{49} X_{52} X_{57} X_{60} X_{63}$} \\ 
\hline
$K^X_{16}$ & \multicolumn{3}{|l|}{$X_{9} X_{15} X_{28} X_{42} X_{48} X_{49} X_{51} X_{62}$} \\
\hline
\end{tabular}
\vspace{10pt}
\caption{Stabilizer generators of the distance-5 3D color code.}
\label{Stabilizers_3D_d5}
\end{table*}

\begin{table*}[htbp]
\centering
\begin{tabular}{| p{1.5cm} | p{6cm} | p{1.5cm} | p{6cm} |}
\hline
\multicolumn{4}{|c|}{$X$-type Stabilizers} \\
\hline
$S^X_{1}$ & $X_{0} X_{1} X_{2} X_{3}$ & $S^X_{2}$ & $X_{1} X_{3} X_{5} X_{6}$ \\
\hline
$S^X_{3}$ & $X_{5} X_{6} X_{8} X_{9} X_{12} X_{13}$ & $S^X_{4}$ & $X_{9} X_{13} X_{17} X_{18}$ \\
\hline
$S^X_{5}$ & $X_{2} X_{3} X_{4} X_{5} X_{7} X_{8}$ & $S^X_{6}$ & $X_{4} X_{7} X_{10} X_{11}$ \\
\hline
$S^X_{7}$ & $X_{10} X_{11} X_{14} X_{15}$ & $S^X_{8}$ & $X_{7} X_{8} X_{11} X_{12} X_{15} X_{16}$ \\
\hline
$S^X_{9}$ & $X_{12} X_{13} X_{16} X_{17}$ & & \\
\hline
\multicolumn{4}{|c|}{$Z$-type Stabilizers} \\
\hline
$S^Z_{1}$ & $Z_{0} Z_{1} Z_{2} Z_{3}$ & $S^Z_{2}$ & $Z_{1} Z_{3} Z_{5} Z_{6}$ \\
\hline
$S^Z_{3}$ & $Z_{5} Z_{6} Z_{8} Z_{9} Z_{12} Z_{13}$ & $S^Z_{4}$ & $Z_{9} Z_{13} Z_{17} Z_{18}$ \\
\hline
$S^Z_{5}$ & $Z_{2} Z_{3} Z_{4} Z_{5} Z_{7} Z_{8}$ & $S^Z_{6}$ & $Z_{4} Z_{7} Z_{10} Z_{11}$ \\
\hline
$S^Z_{7}$ & $Z_{10} Z_{11} Z_{14} Z_{15}$ & $S^Z_{8}$ & $Z_{7} Z_{8} Z_{11} Z_{12} Z_{15} Z_{16}$ \\
\hline
$S^Z_{9}$ & $Z_{12} Z_{13} Z_{16} Z_{17}$ & & \\
\hline
\end{tabular}
\vspace{10pt}
\caption{Stabilizer generators of the distance-5 2D color code.}
\label{Stabilizers_2D_d5}
\end{table*}

\begin{table*}[htbp]
\centering
\begin{tabular}{| p{1cm} | p{4cm} |p{1cm} | p{4cm} |p{1cm} | p{4cm} |}
\hline
\multicolumn{6}{|c|}{$Z$-type Stabilizers} \\
\hline
$K^Z_{1}$ & $Z_{43} Z_{68} Z_{100} Z_{157}$ & $K^Z_{2}$ & $Z_{98} Z_{100} Z_{133} Z_{157}$ & $K^Z_{3}$ & $Z_{69} Z_{71} Z_{98} Z_{102} Z_{118} Z_{133}$ \\
\hline
$K^Z_{4}$ & $Z_{26} Z_{69} Z_{112} Z_{118}$ & $K^Z_{5}$ & $Z_{8} Z_{12} Z_{13} Z_{26} Z_{92} Z_{112}$ & $K^Z_{6}$ & $Z_{8} Z_{14} Z_{92} Z_{109}$ \\
\hline
$K^Z_{7}$ & $Z_{12} Z_{13} Z_{81} Z_{95} Z_{96} Z_{174}$ & $K^Z_{8}$ & $Z_{13} Z_{69} Z_{81} Z_{102} Z_{112} Z_{126}$ & $K^Z_{9}$ & $Z_{5} Z_{81} Z_{95} Z_{126} Z_{132} Z_{171}$ \\
\hline
$K^Z_{10}$ & $Z_{59} Z_{71} Z_{102} Z_{126} Z_{167} Z_{171}$ & $K^Z_{11}$ & $Z_{5} Z_{54} Z_{62} Z_{88} Z_{132} Z_{173}$ & $K^Z_{12}$ & $Z_{59} Z_{68} Z_{71} Z_{133} Z_{138} Z_{157}$ \\
\hline
$K^Z_{13}$ & $Z_{5} Z_{54} Z_{77} Z_{167} Z_{168} Z_{171}$ & $K^Z_{14}$ & $Z_{8} Z_{12} Z_{14} Z_{96}$ & $K^Z_{15}$ & $Z_{88} Z_{95} Z_{132} Z_{174}$ \\
\hline
$K^Z_{16}$ & $Z_{57} Z_{62} Z_{64} Z_{173}$ & $K^Z_{17}$ & $Z_{59} Z_{77} Z_{138} Z_{167}$ & $K^Z_{18}$ & $Z_{54} Z_{57} Z_{62} Z_{168}$ \\
\hline
$K^Z_{19}$ & $Z_{22} Z_{57} Z_{62} Z_{87}$ & $K^Z_{20}$ & $Z_{26} Z_{27} Z_{60} Z_{112}$ & $K^Z_{21}$ & $Z_{39} Z_{81} Z_{95} Z_{124}$ \\
\hline
$K^Z_{22}$ & $Z_{13} Z_{42} Z_{78} Z_{81} Z_{116} Z_{124}$ & $K^Z_{23}$ & $Z_{41} Z_{45} Z_{55} Z_{125} Z_{167} Z_{171}$ & $K^Z_{24}$ & $Z_{8} Z_{14} Z_{29} Z_{105}$ \\
\hline
$K^Z_{25}$ & $Z_{54} Z_{131} Z_{134} Z_{168}$ & $K^Z_{26}$ & $Z_{56} Z_{69} Z_{102} Z_{147}$ & $K^Z_{27}$ & $Z_{56} Z_{60} Z_{69} Z_{93} Z_{112} Z_{144}$ \\
\hline
$K^Z_{28}$ & $Z_{71} Z_{133} Z_{146} Z_{152}$ & $K^Z_{29}$ & $Z_{1} Z_{8} Z_{29} Z_{92}$ & $K^Z_{30}$ & $Z_{13} Z_{60} Z_{78} Z_{112}$ \\
\hline
$K^Z_{31}$ & $Z_{81} Z_{124} Z_{126} Z_{140}$ & $K^Z_{32}$ & $Z_{12} Z_{13} Z_{48} Z_{78}$ & $K^Z_{33}$ & $Z_{8} Z_{12} Z_{29} Z_{48}$ \\
\hline
$K^Z_{34}$ & $Z_{28} Z_{35} Z_{68} Z_{157}$ & $K^Z_{35}$ & $Z_{5} Z_{15} Z_{19} Z_{54} Z_{101} Z_{131}$ & $K^Z_{36}$ & $Z_{22} Z_{62} Z_{172} Z_{173}$ \\
\hline
$K^Z_{37}$ & $Z_{5} Z_{19} Z_{125} Z_{171}$ & $K^Z_{38}$ & $Z_{58} Z_{98} Z_{133} Z_{146}$ & $K^Z_{39}$ & $Z_{28} Z_{61} Z_{133} Z_{136} Z_{146} Z_{157}$ \\
\hline
$K^Z_{40}$ & $Z_{22} Z_{54} Z_{62} Z_{131}$ & $K^Z_{41}$ & $Z_{32} Z_{59} Z_{137} Z_{138}$ & $K^Z_{42}$ & $Z_{20} Z_{32} Z_{59} Z_{71} Z_{129} Z_{152}$ \\
\hline
$K^Z_{43}$ & $Z_{32} Z_{45} Z_{59} Z_{167}$ & $K^Z_{44}$ & $Z_{45} Z_{72} Z_{77} Z_{167}$ & $K^Z_{45}$ & $Z_{28} Z_{100} Z_{154} Z_{157}$ \\
\hline
$K^Z_{46}$ & $Z_{56} Z_{69} Z_{114} Z_{118}$ & $K^Z_{47}$ & $Z_{71} Z_{102} Z_{147} Z_{152}$ & $K^Z_{48}$ & $Z_{12} Z_{21} Z_{48} Z_{96}$ \\
\hline
$K^Z_{49}$ & $Z_{7} Z_{39} Z_{95} Z_{174}$ & $K^Z_{50}$ & $Z_{39} Z_{95} Z_{132} Z_{162}$ & $K^Z_{51}$ & $Z_{88} Z_{122} Z_{132} Z_{162}$ \\
\hline
$K^Z_{52}$ & $Z_{5} Z_{19} Z_{132} Z_{162}$ & $K^Z_{53}$ & $Z_{30} Z_{91} Z_{102} Z_{126} Z_{140} Z_{147}$ & $K^Z_{54}$ & $Z_{0} Z_{38} Z_{46} Z_{85} Z_{111} Z_{113}$ \\
\hline
$K^Z_{55}$ & $Z_{41} Z_{49} Z_{84} Z_{151}$ & $K^Z_{56}$ & $Z_{91} Z_{116} Z_{124} Z_{140}$ & $K^Z_{57}$ & $Z_{2} Z_{6} Z_{65} Z_{117}$ \\
\hline
$K^Z_{58}$ & $Z_{58} Z_{61} Z_{110} Z_{146}$ & $K^Z_{59}$ & $Z_{31} Z_{42} Z_{89} Z_{116}$ & $K^Z_{60}$ & $Z_{15} Z_{41} Z_{55} Z_{101} Z_{151} Z_{163}$ \\
\hline
$K^Z_{61}$ & $Z_{18} Z_{33} Z_{53} Z_{106} Z_{160} Z_{164}$ & $K^Z_{62}$ & $Z_{73} Z_{79} Z_{86} Z_{89}$ & $K^Z_{63}$ & $Z_{25} Z_{86} Z_{89} Z_{91} Z_{116} Z_{120}$ \\
\hline
$K^Z_{64}$ & $Z_{36} Z_{51} Z_{63} Z_{74}$ & $K^Z_{65}$ & $Z_{3} Z_{33} Z_{106} Z_{161}$ & $K^Z_{66}$ & $Z_{10} Z_{19} Z_{101} Z_{103} Z_{122} Z_{162}$ \\
\hline
$K^Z_{67}$ & $Z_{37} Z_{50} Z_{66} Z_{70} Z_{110} Z_{139}$ & $K^Z_{68}$ & $Z_{4} Z_{16} Z_{18} Z_{53} Z_{84} Z_{120}$ & $K^Z_{69}$ & $Z_{18} Z_{20} Z_{164} Z_{169}$ \\
\hline
$K^Z_{70}$ & $Z_{121} Z_{135} Z_{148} Z_{153}$ & $K^Z_{71}$ & $Z_{11} Z_{75} Z_{143} Z_{170}$ & $K^Z_{72}$ & $Z_{10} Z_{17} Z_{83} Z_{103}$ \\
\hline
$K^Z_{73}$ & $Z_{11} Z_{47} Z_{67} Z_{75}$ & $K^Z_{74}$ & $Z_{76} Z_{80} Z_{82} Z_{108} Z_{127} Z_{150}$ & $K^Z_{75}$ & $Z_{53} Z_{61} Z_{70} Z_{110} Z_{129} Z_{160}$ \\
\hline
$K^Z_{76}$ & $Z_{15} Z_{131} Z_{134} Z_{163}$ & $K^Z_{77}$ & $Z_{28} Z_{136} Z_{139} Z_{154}$ & $K^Z_{78}$ & $Z_{6} Z_{16} Z_{49} Z_{84} Z_{117} Z_{130}$ \\
\hline
$K^Z_{79}$ & $Z_{111} Z_{121} Z_{135} Z_{142}$ & $K^Z_{80}$ & $Z_{111} Z_{113} Z_{135} Z_{153}$ & $K^Z_{81}$ & $Z_{3} Z_{33} Z_{34} Z_{46} Z_{149} Z_{160}$ \\
\hline
$K^Z_{82}$ & $Z_{16} Z_{41} Z_{55} Z_{84}$ & $K^Z_{83}$ & $Z_{18} Z_{49} Z_{84} Z_{164}$ & $K^Z_{84}$ & $Z_{40} Z_{50} Z_{139} Z_{154}$ \\
\hline
$K^Z_{85}$ & $Z_{22} Z_{87} Z_{158} Z_{172}$ & $K^Z_{86}$ & $Z_{41} Z_{45} Z_{72} Z_{151}$ & $K^Z_{87}$ & $Z_{3} Z_{108} Z_{127} Z_{161}$ \\
\hline
$K^Z_{88}$ & $Z_{10} Z_{52} Z_{83} Z_{97} Z_{107} Z_{115}$ & $K^Z_{89}$ & $Z_{7} Z_{39} Z_{79} Z_{89} Z_{116} Z_{124}$ & $K^Z_{90}$ & $Z_{47} Z_{80} Z_{82} Z_{94}$ \\
\hline
$K^Z_{91}$ & $Z_{9} Z_{17} Z_{83} Z_{107}$ & $K^Z_{92}$ & $Z_{37} Z_{66} Z_{108} Z_{127}$ & $K^Z_{93}$ & $Z_{47} Z_{67} Z_{82} Z_{155}$ \\
\hline
$K^Z_{94}$ & $Z_{15} Z_{83} Z_{107} Z_{163}$ & $K^Z_{95}$ & $Z_{11} Z_{47} Z_{82} Z_{150}$ & $K^Z_{96}$ & $Z_{1} Z_{29} Z_{105} Z_{119}$ \\
\hline
$K^Z_{97}$ & $Z_{25} Z_{65} Z_{85} Z_{99} Z_{117} Z_{130}$ & $K^Z_{98}$ & $Z_{11} Z_{75} Z_{150} Z_{166}$ & $K^Z_{99}$ & $Z_{80} Z_{82} Z_{155} Z_{156}$ \\
\hline
$K^Z_{100}$ & $Z_{3} Z_{76} Z_{108} Z_{149}$ & $K^Z_{101}$ & $Z_{21} Z_{31} Z_{42} Z_{48} Z_{78} Z_{104}$ & $K^Z_{102}$ & $Z_{10} Z_{15} Z_{83} Z_{101}$ \\
\hline
$K^Z_{103}$ & $Z_{25} Z_{38} Z_{85} Z_{86}$ & $K^Z_{104}$ & $Z_{111} Z_{113} Z_{142} Z_{165}$ & $K^Z_{105}$ & $Z_{24} Z_{44} Z_{97} Z_{115}$ \\
\hline
$K^Z_{106}$ & $Z_{61} Z_{129} Z_{146} Z_{152}$ & $K^Z_{107}$ & $Z_{6} Z_{52} Z_{97} Z_{117}$ & $K^Z_{108}$ & $Z_{42} Z_{60} Z_{78} Z_{144}$ \\
\hline
$K^Z_{109}$ & $Z_{23} Z_{38} Z_{73} Z_{86}$ & $K^Z_{110}$ & $Z_{31} Z_{104} Z_{121} Z_{135}$ & $K^Z_{111}$ & $Z_{24} Z_{52} Z_{97} Z_{123}$ \\
\hline
$K^Z_{112}$ & $Z_{0} Z_{37} Z_{70} Z_{76} Z_{108} Z_{159}$ & $K^Z_{113}$ & $Z_{34} Z_{70} Z_{159} Z_{160}$ & $K^Z_{114}$ & $Z_{25} Z_{44} Z_{73} Z_{86} Z_{115} Z_{130}$ \\
\hline
$K^Z_{115}$ & $Z_{27} Z_{60} Z_{144} Z_{145}$ & $K^Z_{116}$ & $Z_{4} Z_{34} Z_{53} Z_{160}$ & $K^Z_{117}$ & $Z_{4} Z_{30} Z_{53} Z_{129} Z_{147} Z_{152}$ \\
\hline
$K^Z_{118}$ & $Z_{2} Z_{36} Z_{65} Z_{74}$ & $K^Z_{119}$ & $Z_{51} Z_{74} Z_{143} Z_{170}$ & $K^Z_{120}$ & $Z_{4} Z_{30} Z_{34} Z_{93} Z_{141} Z_{159}$ \\
\hline
$K^Z_{121}$ & $Z_{31} Z_{42} Z_{135} Z_{144} Z_{145} Z_{153}$ & $K^Z_{122}$ & $Z_{18} Z_{20} Z_{53} Z_{129}$ & $K^Z_{123}$ & $Z_{97} Z_{115} Z_{117} Z_{130}$ \\
\hline
$K^Z_{124}$ & $Z_{23} Z_{38} Z_{85} Z_{99} Z_{143} Z_{170}$ & $K^Z_{125}$ & $Z_{9} Z_{17} Z_{24} Z_{44} Z_{103} Z_{123}$ & $K^Z_{126}$ & $Z_{0} Z_{46} Z_{76} Z_{149}$ \\
\hline
$K^Z_{127}$ & $Z_{65} Z_{74} Z_{99} Z_{170}$ & $K^Z_{128}$ & $Z_{46} Z_{85} Z_{99} Z_{149}$ & $K^Z_{129}$ & $Z_{51} Z_{63} Z_{67} Z_{75} Z_{128} Z_{143}$ \\
\hline
$K^Z_{130}$ & $Z_{67} Z_{128} Z_{155} Z_{156}$ & $K^Z_{131}$ & $Z_{50} Z_{66} Z_{90} Z_{106} Z_{164} Z_{169}$ & $K^Z_{132}$ & $Z_{2} Z_{36} Z_{80} Z_{94} Z_{127} Z_{161}$ \\
\hline
$K^Z_{133}$ & $Z_{2} Z_{6} Z_{49} Z_{106} Z_{161} Z_{164}$ & $K^Z_{134}$ & $Z_{0} Z_{113} Z_{141} Z_{145} Z_{153} Z_{159}$ &  &  \\
\hline
\end{tabular}
\vspace{10pt}
\caption{$Z$-type stabilizer generators of the distance-7 3D color code.}
\label{Stabilizers_3D_d7_kx_z}
\end{table*}

\begin{table*}[htbp]
\centering
\begin{tabular}{| p{1.5cm} | p{14cm} |}
\hline
\multicolumn{2}{|c|}{$X$-type Stabilizers} \\
\hline
$K^X_{1}$ & $X_{28} X_{35} X_{40} X_{43} X_{68} X_{100} X_{154} X_{157}$ \\
\hline
$K^X_{2}$ & $X_{28} X_{58} X_{61} X_{98} X_{100} X_{110} X_{133} X_{136} X_{139} X_{146} X_{154} X_{157}$ \\
\hline
$K^X_{3}$ & $X_{56} X_{58} X_{69} X_{71} X_{98} X_{102} X_{114} X_{118} X_{133} X_{146} X_{147} X_{152}$ \\
\hline
$K^X_{4}$ & $X_{26} X_{27} X_{56} X_{60} X_{69} X_{93} X_{112} X_{114} X_{118} X_{141} X_{144} X_{145}$ \\
\hline
$K^X_{5}$ & $X_{1} X_{8} X_{12} X_{13} X_{26} X_{27} X_{29} X_{48} X_{60} X_{78} X_{92} X_{112}$ \\
\hline
$K^X_{6}$ & $X_{1} X_{8} X_{14} X_{29} X_{92} X_{105} X_{109} X_{119}$ \\
\hline
$K^X_{7}$ & $X_{7} X_{12} X_{13} X_{21} X_{31} X_{39} X_{42} X_{48} X_{78} X_{79} X_{81} X_{89} X_{95} X_{96} X_{104} X_{116} X_{124} X_{174}$ \\
\hline
$K^X_{8}$ & $X_{13} X_{30} X_{42} X_{56} X_{60} X_{69} X_{78} X_{81} X_{91} X_{93} X_{102} X_{112} X_{116} X_{124} X_{126} X_{140} X_{144} X_{147}$ \\
\hline
$K^X_{9}$ & $X_{5} X_{19} X_{39} X_{81} X_{95} X_{124} X_{125} X_{126} X_{132} X_{140} X_{162} X_{171}$ \\
\hline
$K^X_{10}$ & $X_{4} X_{16} X_{18} X_{20} X_{30} X_{32} X_{41} X_{45} X_{53} X_{55} X_{59} X_{71} X_{84} X_{91} X_{102} X_{120} X_{125} X_{126} X_{129} X_{140} X_{147} X_{152} X_{167} X_{171}$ \\
\hline
$K^X_{11}$ & $X_{5} X_{10} X_{15} X_{17} X_{19} X_{22} X_{54} X_{62} X_{83} X_{88} X_{101} X_{103} X_{122} X_{131} X_{132} X_{162} X_{172} X_{173}$ \\
\hline
$K^X_{12}$ & $X_{20} X_{28} X_{32} X_{35} X_{59} X_{61} X_{68} X_{71} X_{90} X_{129} X_{133} X_{136} X_{137} X_{138} X_{146} X_{152} X_{157} X_{169}$ \\
\hline
$K^X_{13}$ & $X_{5} X_{15} X_{19} X_{41} X_{45} X_{54} X_{55} X_{72} X_{77} X_{101} X_{125} X_{131} X_{134} X_{151} X_{163} X_{167} X_{168} X_{171}$ \\
\hline
$K^X_{14}$ & $X_{8} X_{12} X_{14} X_{21} X_{29} X_{48} X_{96} X_{105}$ \\
\hline
$K^X_{15}$ & $X_{7} X_{39} X_{88} X_{95} X_{122} X_{132} X_{162} X_{174}$ \\
\hline
$K^X_{16}$ & $X_{22} X_{57} X_{62} X_{64} X_{87} X_{158} X_{172} X_{173}$ \\
\hline
$K^X_{17}$ & $X_{32} X_{45} X_{59} X_{72} X_{77} X_{137} X_{138} X_{167}$ \\
\hline
$K^X_{18}$ & $X_{22} X_{54} X_{57} X_{62} X_{87} X_{131} X_{134} X_{168}$ \\
\hline
$K^X_{19}$ & $X_{18} X_{20} X_{33} X_{37} X_{50} X_{53} X_{61} X_{66} X_{70} X_{90} X_{106} X_{110} X_{129} X_{136} X_{139} X_{160} X_{164} X_{169}$ \\
\hline
$K^X_{20}$ & $X_{6} X_{10} X_{15} X_{16} X_{41} X_{49} X_{52} X_{55} X_{83} X_{84} X_{97} X_{101} X_{107} X_{115} X_{117} X_{130} X_{151} X_{163}$ \\
\hline
$K^X_{21}$ & $X_{2} X_{3} X_{11} X_{36} X_{47} X_{65} X_{74} X_{76} X_{80} X_{82} X_{94} X_{99} X_{108} X_{127} X_{149} X_{150} X_{161} X_{170}$ \\
\hline
$K^X_{22}$ & $X_{2} X_{3} X_{4} X_{6} X_{16} X_{18} X_{25} X_{33} X_{34} X_{46} X_{49} X_{53} X_{65} X_{84} X_{85} X_{99} X_{106} X_{117} X_{120} X_{130} X_{149} X_{160} X_{161} X_{164}$ \\
\hline
$K^X_{23}$ & $X_{9} X_{10} X_{17} X_{24} X_{44} X_{52} X_{83} X_{97} X_{103} X_{107} X_{115} X_{123}$ \\
\hline
$K^X_{24}$ & $X_{11} X_{36} X_{47} X_{51} X_{63} X_{67} X_{74} X_{75} X_{94} X_{128} X_{143} X_{170}$ \\
\hline
$K^X_{25}$ & $X_{28} X_{35} X_{40} X_{50} X_{90} X_{136} X_{139} X_{154}$ \\
\hline
$K^X_{26}$ & $X_{18} X_{20} X_{32} X_{41} X_{45} X_{49} X_{72} X_{84} X_{137} X_{151} X_{164} X_{169}$ \\
\hline
$K^X_{27}$ & $X_{3} X_{33} X_{37} X_{66} X_{106} X_{108} X_{127} X_{161}$ \\
\hline
$K^X_{28}$ & $X_{9} X_{15} X_{17} X_{22} X_{83} X_{87} X_{107} X_{131} X_{134} X_{158} X_{163} X_{172}$ \\
\hline
$K^X_{29}$ & $X_{2} X_{6} X_{24} X_{36} X_{51} X_{52} X_{63} X_{65} X_{74} X_{97} X_{117} X_{123}$ \\
\hline
$K^X_{30}$ & $X_{47} X_{67} X_{80} X_{82} X_{94} X_{128} X_{155} X_{156}$ \\
\hline
$K^X_{31}$ & $X_{0} X_{3} X_{33} X_{34} X_{37} X_{46} X_{70} X_{76} X_{108} X_{149} X_{159} X_{160}$ \\
\hline
$K^X_{32}$ & $X_{23} X_{24} X_{25} X_{38} X_{44} X_{51} X_{65} X_{73} X_{74} X_{85} X_{86} X_{97} X_{99} X_{115} X_{117} X_{130} X_{143} X_{170}$ \\
\hline
$K^X_{33}$ & $X_{11} X_{47} X_{67} X_{75} X_{82} X_{150} X_{155} X_{166}$ \\
\hline
$K^X_{34}$ & $X_{0} X_{4} X_{25} X_{30} X_{31} X_{34} X_{38} X_{42} X_{46} X_{85} X_{86} X_{89} X_{91} X_{93} X_{111} X_{113} X_{116} X_{120} X_{135} X_{141} X_{144} X_{145} X_{153} X_{159}$ \\
\hline
$K^X_{35}$ & $X_{23} X_{31} X_{38} X_{73} X_{79} X_{86} X_{89} X_{104} X_{111} X_{121} X_{135} X_{142}$ \\
\hline
$K^X_{36}$ & $X_{4} X_{30} X_{34} X_{53} X_{56} X_{58} X_{61} X_{70} X_{93} X_{110} X_{114} X_{129} X_{141} X_{146} X_{147} X_{152} X_{159} X_{160}$ \\
\hline
$K^X_{37}$ & $X_{7} X_{10} X_{16} X_{19} X_{25} X_{39} X_{44} X_{55} X_{73} X_{79} X_{86} X_{89} X_{91} X_{101} X_{103} X_{115} X_{116} X_{120} X_{122} X_{124} X_{125} X_{130} X_{140} X_{162}$ \\
\hline
$K^X_{38}$ & $X_{0} X_{11} X_{23} X_{38} X_{46} X_{75} X_{76} X_{85} X_{99} X_{111} X_{113} X_{142} X_{143} X_{149} X_{150} X_{165} X_{166} X_{170}$ \\
\hline
$K^X_{39}$ & $X_{111} X_{113} X_{121} X_{135} X_{142} X_{148} X_{153} X_{165}$ \\
\hline
$K^X_{40}$ & $X_{1} X_{21} X_{27} X_{29} X_{31} X_{42} X_{48} X_{60} X_{78} X_{104} X_{105} X_{119} X_{121} X_{135} X_{144} X_{145} X_{148} X_{153}$ \\
\hline
\end{tabular}
\vspace{10pt}
\caption{$X$-type stabilizer generators of the distance-7 3D color code.}
\label{Stabilizers_3D_d7_kz_x}
\end{table*}

\begin{table*}[htbp]
\centering
\begin{tabular}{| p{1.5cm} | p{6cm} | p{1.5cm} | p{6cm} |}
\hline
\multicolumn{4}{|c|}{$X$-type Stabilizers from $s_x$} \\
\hline
$S^X_{1}$ & $X_{24} X_{25} X_{30} X_{31}$ & $S^X_{2}$ & $X_{14} X_{19} X_{24} X_{25}$ \\
\hline
$S^X_{3}$ & $X_{10} X_{11} X_{14} X_{15} X_{19} X_{20}$ & $S^X_{4}$ & $X_{4} X_{7} X_{10} X_{11}$ \\
\hline
$S^X_{5}$ & $X_{2} X_{3} X_{4} X_{5} X_{7} X_{8}$ & $S^X_{6}$ & $X_{0} X_{1} X_{2} X_{3}$ \\
\hline
$S^X_{7}$ & $X_{5} X_{6} X_{8} X_{9} X_{12} X_{13}$ & $S^X_{8}$ & $X_{7} X_{8} X_{11} X_{12} X_{15} X_{16}$ \\
\hline
$S^X_{9}$ & $X_{12} X_{13} X_{16} X_{17} X_{21} X_{22}$ & $S^X_{10}$ & $X_{15} X_{16} X_{20} X_{21} X_{26} X_{27}$ \\
\hline
$S^X_{11}$ & $X_{17} X_{18} X_{22} X_{23} X_{28} X_{29}$ & $S^X_{12}$ & $X_{19} X_{20} X_{25} X_{26} X_{31} X_{32}$ \\
\hline
$S^X_{13}$ & $X_{21} X_{22} X_{27} X_{28} X_{33} X_{34}$ & $S^X_{14}$ & $X_{1} X_{3} X_{5} X_{6}$ \\
\hline
$S^X_{15}$ & $X_{9} X_{13} X_{17} X_{18}$ & $S^X_{16}$ & $X_{23} X_{29} X_{35} X_{36}$ \\
\hline
$S^X_{17}$ & $X_{26} X_{27} X_{32} X_{33}$ & $S^X_{18}$ & $X_{28} X_{29} X_{34} X_{35}$ \\
\hline
\multicolumn{4}{|c|}{$Z$-type Stabilizers from $s_z$} \\
\hline
$S^Z_{1}$ & $Z_{24} Z_{25} Z_{30} Z_{31}$ & $S^Z_{2}$ & $Z_{14} Z_{19} Z_{24} Z_{25}$ \\
\hline
$S^Z_{3}$ & $Z_{10} Z_{11} Z_{14} Z_{15} Z_{19} Z_{20}$ & $S^Z_{4}$ & $Z_{4} Z_{7} Z_{10} Z_{11}$ \\
\hline
$S^Z_{5}$ & $Z_{2} Z_{3} Z_{4} Z_{5} Z_{7} Z_{8}$ & $S^Z_{6}$ & $Z_{0} Z_{1} Z_{2} Z_{3}$ \\
\hline
$S^Z_{7}$ & $Z_{5} Z_{6} Z_{8} Z_{9} Z_{12} Z_{13}$ & $S^Z_{8}$ & $Z_{7} Z_{8} Z_{11} Z_{12} Z_{15} Z_{16}$ \\
\hline
$S^Z_{9}$ & $Z_{12} Z_{13} Z_{16} Z_{17} Z_{21} Z_{22}$ & $S^Z_{10}$ & $Z_{15} Z_{16} Z_{20} Z_{21} Z_{26} Z_{27}$ \\
\hline
$S^Z_{11}$ & $Z_{17} Z_{18} Z_{22} Z_{23} Z_{28} Z_{29}$ & $S^Z_{12}$ & $Z_{19} Z_{20} Z_{25} Z_{26} Z_{31} Z_{32}$ \\
\hline
$S^Z_{13}$ & $Z_{21} Z_{22} Z_{27} Z_{28} Z_{33} Z_{34}$ & $S^Z_{14}$ & $Z_{1} Z_{3} Z_{5} Z_{6}$ \\
\hline
$S^Z_{15}$ & $Z_{9} Z_{13} Z_{17} Z_{18}$ & $S^Z_{16}$ & $Z_{23} Z_{29} Z_{35} Z_{36}$ \\
\hline
$S^Z_{17}$ & $Z_{26} Z_{27} Z_{32} Z_{33}$ & $S^Z_{18}$ & $Z_{28} Z_{29} Z_{34} Z_{35}$ \\
\hline
\end{tabular}
\vspace{10pt}
\caption{Stabilizer generators of the distance-7 2D color code.}
\label{Stabilizers_2D_d7}
\end{table*}

\begin{table*}[htbp]
\centering
\begin{tabular}{| p{1.5cm} | p{14cm} |}
\hline
\multicolumn{2}{|c|}{Flag sequences used for $d=3$} \\
\hline
Weight & Sequence \\
\hline
Weight 3 &
$F_1 D_1 D_2 F_2 D_3 F_1 F_2$ \\
\hline
Weight 4 &
$F_1 D_1 D_2 F_2 D_3 D_4 F_1 F_2$ \\
\hline
Weight 5 &
$F_1 D_1 F_2 D_2 F_3 D_3 F_4 D_4 F_5 D_5 F_2 F_4 F_1 F_5 F_3$ \\
\hline
Weight 8 &
$F_1 D_1 F_2 D_2 F_3 D_3 F_4 D_4 F_5 D_5 F_2 D_6 F_6 D_7 F_3 D_8 F_1 F_4 F_5 F_6$ \\
\hline
Weight 9 &
$F_1 D_1 F_2 D_2 F_3 D_3 F_4 D_4 F_5 D_5 F_2 D_6 F_6 D_7 F_3 D_8 F_5 D_9 F_1 F_6 F_4$ \\
\hline
\end{tabular}
\vspace{10pt}
\caption{Flag sequences used for the distance-3 circuit-level interface implementation.}
\label{Flag_sequences_d3}
\end{table*}

\begin{table*}[htbp]
\centering
\begin{tabular}{| p{1.5cm} | p{14cm} |}
\hline
\multicolumn{2}{|c|}{Flag sequences used for $d=5$} \\
\hline
Weight & Sequence \\
\hline
Weight 3 &
$F_1 D_1 D_2 F_2 D_3 F_1 F_2$ \\
\hline
Weight 4 &
$F_1 D_1 D_2 F_2 D_3 D_4 F_1 F_2$ \\
\hline
Weight 5 &
$F_1 D_1 F_2 D_2 F_3 D_3 F_4 D_4 F_5 D_5 F_2 F_4 F_1 F_5 F_3$ \\
\hline
Weight 6 &
$F_1 F_2 F_3 D_1 F_4 D_2 F_5 D_3 F_1 D_4 F_6 D_5 F_2 D_6 F_3 F_4 F_5 F_6$ \\
\hline
Weight 7 &
$F_1 D_1 F_2 D_2 F_3 D_3 F_4 D_4 F_5 D_5 F_2 D_6 F_6 D_7 F_3 F_5 F_1 F_6 F_4$ \\
\hline
Weight 8 &
$F_1 D_1 F_2 D_2 F_3 D_3 F_4 D_4 F_5 D_5 F_2 D_6 F_6 D_7 F_3 D_8 F_1 F_4 F_5 F_6$ \\
\hline
Weight 9 &
$F_1 D_1 F_2 D_2 F_3 D_3 F_4 D_4 F_5 D_5 F_2 D_6 F_6 D_7 F_3 D_8 F_5 D_9 F_1 F_6 F_4$ \\
\hline
Weight 12 &
$F_1 D_1 F_2 D_2 F_3 D_3 F_4 D_4 F_5 D_5 F_2 D_6 F_6 D_7 F_3 D_8 F_7 D_9 F_4 D_{10} F_8 D_{11} F_1 D_{12} F_5 F_6 F_7 F_8$ \\
\hline
Weight 13 &
$F_1 D_1 F_2 D_2 F_3 D_3 F_4 D_4 F_5 D_5 F_2 D_6 F_6 D_7 F_3 D_8 F_7 D_9 F_4 D_{10} F_8 D_{11} F_5 D_{12} F_7 D_{13} F_1 F_8 F_6$ \\
\hline
Weight 18 &
$\begin{array}{@{}lc@{}}
F_1 D_1 F_2 D_2 F_3 D_3 F_4 D_4 F_5 D_5 F_2 D_6 F_6 D_7 F_3 D_8 F_7 D_9 F_4 D_{10} F_8 D_{11}\\ F_5 D_{12} F_9 D_{13} F_6 D_{14} F_{10} D_{15} F_7 D_{16} F_{11} D_{17} F_1 D_{18} F_8 F_9 F_{10} F_{11}
\end{array}$ \\
\hline
Weight 19 &
$\begin{array}{@{}lc@{}}
F_1 D_1 F_2 D_2 F_3 D_3 F_4 D_4 F_5 D_5 F_2 D_6 F_6 D_7 F_3 D_8 F_7 D_9 F_4 D_{10} F_8 D_{11}\\ F_5 D_{12} F_9 D_{13} F_6 D_{14} F_{10} D_{15} F_7 D_{16} F_{11} D_{17} F_8 D_{18} F_{10} D_{19} F_1 F_{11} F_9
\end{array}$ \\
\hline
\end{tabular}
\vspace{10pt}
\caption{Flag sequences used for the distance-5 circuit-level interface implementation.}
\label{Flag_sequences_d5}
\end{table*}

\begin{table*}[htbp]
\centering
\begin{tabular}{| p{1.5cm} | p{14cm} |}
\hline
\multicolumn{2}{|c|}{Flag sequences used for $d=7$} \\
\hline
Weight & Sequence \\
\hline
Weight 3 &
$F_1 D_1 D_2 F_2 D_3 F_1 F_2$ \\
\hline
Weight 4 &
$F_1 D_1 D_2 F_2 D_3 D_4 F_1 F_2$ \\
\hline
Weight 5 &
$F_1 D_5 F_2 F_3 F_4 D_4 F_5 F_6 F_7 D_3 F_2 F_4 F_6 D_2 F_1 F_3 F_5 D_1 F_7$ \\
\hline
Weight 6 &
$F_1 D_6 F_2 F_3 F_4 D_5 F_5 F_6 F_7 D_4 F_2 F_4 F_6 D_3 F_1 F_3 F_5 D_2 F_7 D_1$ \\
\hline
Weight 7 &
$F_1 D_7 F_2 F_3 D_6 F_4 F_5 F_6 D_5 F_7 F_2 F_8 D_4 F_3 F_5 F_7 D_3 F_1 F_4 F_6 D_2 F_8 D_1$ \\
\hline
Weight 8 &
$F_1 D_8 F_2 F_3 D_7 F_4 F_5 F_6 D_6 F_7 F_2 F_8 D_5 F_4 F_9 F_3 D_4 F_6 F_8 F_1 D_3 F_5 F_7 D_2 F_9 D_1$ \\
\hline
Weight 9 &
$F_1 D_9 F_2 F_3 D_8 F_4 F_5 D_7 F_6 F_7 F_2 D_6 F_8 F_4 F_9 D_5 F_6 F_{10} F_1 D_4 F_{11} F_5 F_8 D_3 F_{10} F_3 D_2 F_7 F_9 D_1 F_{11}$ \\
\hline
Weight 12 &
$\begin{array}{@{}lc@{}}
F_1 D_{12} F_2 F_3 D_{11} F_4 F_5 D_{10} F_6 F_7 D_9 F_2 F_8 F_4 D_8 F_9 F_6 F_{10} D_7 
\\
F_1 F_{11} F_3 D_6 F_{12} F_5 F_{13} D_5 F_8 F_{10} D_4 F_{12} F_7 D_3 F_9 F_{11} D_2 F_{13} D_1
\end{array}$ \\
\hline
Weight 13 &
$\begin{array}{@{}lc@{}}
F_1 D_{13} F_2 F_3 D_{12} F_4 F_5 D_{11} F_6 F_7 D_{10} F_2 F_8 D_9 F_4 F_9 F_6 D_8 F_{10} F_1 F_{11} 
\\ 
D_7 F_3 F_{12} F_5 D_6 F_{13} F_7 F_{14} D_5 F_9 F_{11} D_4 F_{13} F_8 D_3 F_{10} F_{12} D_2 F_{14} D_1
\end{array}$ \\
\hline
Weight 18 &
$\begin{array}{@{}lc@{}}
F_1 D_{18} F_2 F_3 D_{17} F_4 F_5 D_{16} F_6 F_7 D_{15} F_2 F_8 D_{14} F_4 F_9 D_{13} F_6 F_{10} D_{12} F_1 F_{11} F_3 D_{11} F_{12} F_5 F_{13} D_{10} F_7 F_{14} 
\\
F_9 D_9 F_{15} F_{11} F_{16} D_8 F_{13} F_{17} D_7 F_8 F_{18} D_6 F_{10} F_{19} D_5 F_{14} F_{16} D_4 F_{18} F_{12} D_3 F_{15} F_{17} D_2 F_{19} D_1
\end{array}$ \\
\hline
Weight 19 &
$\begin{array}{@{}lc@{}}
F_1 D_{19} F_2 F_3 D_{18} F_4 F_5 D_{17} F_6 F_7 D_{16} F_2 F_8 D_{15} F_4 F_9 D_{14} F_6 F_{10} D_{13} F_1 F_{11} D_{12} F_3 F_{12} F_5 D_{11} F_{13} F_7 F_{14} 
\\
D_{10} F_9 F_{15} F_{11} D_9 F_{16} F_{13} F_{17} D_8 F_8 F_{18} D_7 F_{10} F_{19} D_6 F_{12} F_{20} D_5 F_{15} F_{17} D_4 F_{19} F_{14} D_3 F_{16} F_{18} D_2 F_{20} D_1
\end{array}$ \\
\hline
Weight 24 &
$\begin{array}{@{}lc@{}}
F_1 D_{24} F_2 F_3 D_{23} F_4 F_5 D_{22} F_6 F_7 D_{21} F_2 F_8 D_{20} F_4 F_9 D_{19} F_6 F_{10} D_{18} F_1 F_{11} D_{17} F_3 F_{12}
\\ 
D_{16} F_5 F_{13} D_{15} F_7 F_{14} F_9 D_{14} F_{15} F_{11} F_{16} D_{13} 
F_{13} F_{17} F_8 D_{12} F_{18} F_{10} F_{19} D_{11} F_{12} F_{20} D_{10}
\\
F_{14} F_{21} D_9 F_{16} F_{22} D_8 F_{18} F_{23} D_7 F_{20} F_{24} D_6 F_{15} F_{25} D_5 F_{19} F_{22} D_4 F_{24} F_{17} D_3 F_{21} F_{23} D_2 F_{25} D_1
\end{array}$ \\
\hline
\end{tabular}
\vspace{10pt}
\caption{Flag sequences used for the distance-7 circuit-level interface implementation.}
\label{Flag_sequences_d7}
\end{table*}

\end{document}